\documentclass[12pt]{iopart}

\newcommand{\thetitle}{A Hydrophobic Gate in an Ion Channel: The
  Closed State of the Nicotinic Acetylcholine Receptor}
\newcommand{\runningtitle}{A Hydrophobic Gate in nAChR}
\newcommand{\thecorrespondingauthor}{To whom correspondence should be
  addressed: Telephone: +44 (0)1865 275371, E-mail:
  \mailto{mark.sansom@bioch.ox.ac.uk}} 
\newcommand{\theauthor}{Oliver Beckstein$^{1,2}$ and Mark S P
  Sansom$^{2}$\footnote[1]{\thecorrespondingauthor}}
\newcommand{\JHU}{$^{1}$ The Johns Hopkins University, School of
  Medicine, Department of Physiology, Biophysics 206, 725 N.{} Wolfe
  St., Baltimore, MD 21205, USA.}
\newcommand{\SBCB}{$^{2}$ Structural Bioinformatics and
  Computational Biochemistry Unit, Department of Biochemistry, University of
  Oxford, South Parks Road, Oxford OX1 3QU, UK}

\usepackage{times}
\usepackage{mathptm}

\usepackage{graphicx}  
\usepackage{amstext}   
\usepackage{url}       
\usepackage{booktabs}

\usepackage[numbers,sort&compress]{natbib}  
\renewcommand{\vec}[1]{\mathbf{#1}}

\newcommand{\NA}{Na$^{+}$}
\newcommand{\CL}{Cl$^{-}$}

\begin{document}

%
\title[\runningtitle]{\thetitle}
\author{\theauthor}
\address{\JHU}\ead{orbeckst@jhmi.edu}
\address{\SBCB}\ead{mark.sansom@bioch.ox.ac.uk}
%

\begin{abstract}
  \noindent%
  The nicotinic acetylcholine receptor (nAChR) is the prototypic
  member of the `Cys-loop' superfamily of ligand-gated ion channels
  which mediate synaptic neurotransmission, and whose other members
  include receptors for glycine, $\gamma$-{}amino\-bu\-ty\-ric acid,
  and serotonin. Cryo-electron microscopy has yielded a three
  dimensional structure of the nAChR in its closed state. However, the
  exact nature and location of the channel gate remains uncertain.
  Although the transmembrane pore is constricted close to its center,
  it is not completely occluded. Rather, the pore has a central
  hydrophobic zone of radius about 3 \AA{}. Model calculations suggest
  that such a constriction may form a hydrophobic gate, preventing
  movement of ions through a channel.  We present a detailed and
  quantitative simulation study of the hydrophobic gating model of the
  nicotinic receptor, in order to fully evaluate this hypothesis. We
  demonstrate that the hydrophobic constriction of the nAChR pore
  indeed forms a closed gate. Potential of mean force (PMF)
  calculations reveal that the constriction presents a barrier of
  height ca.\ $10\,kT$ to the permeation of sodium ions, placing an
  upper bound on the closed channel conductance of $0.3$~pS. Thus, a
  3~\AA{} radius hydrophobic pore can form a functional barrier to the
  permeation of a 1~\AA{} radius Na$^{+}$ ion.  Using a united atom
  force field for the protein instead of an all atom one retains the
  qualitative features but results in differing conductances, showing
  that the PMF is sensitive to the detailed molecular interactions.

\end{abstract}



\submitto{Physical Biology}
\maketitle

\clearpage

\section{Introduction}
\label{sec:introduction}

\begin{figure}[tb]
    \centering
    \hspace*{-10mm}
    \includegraphics[clip,width=65mm]{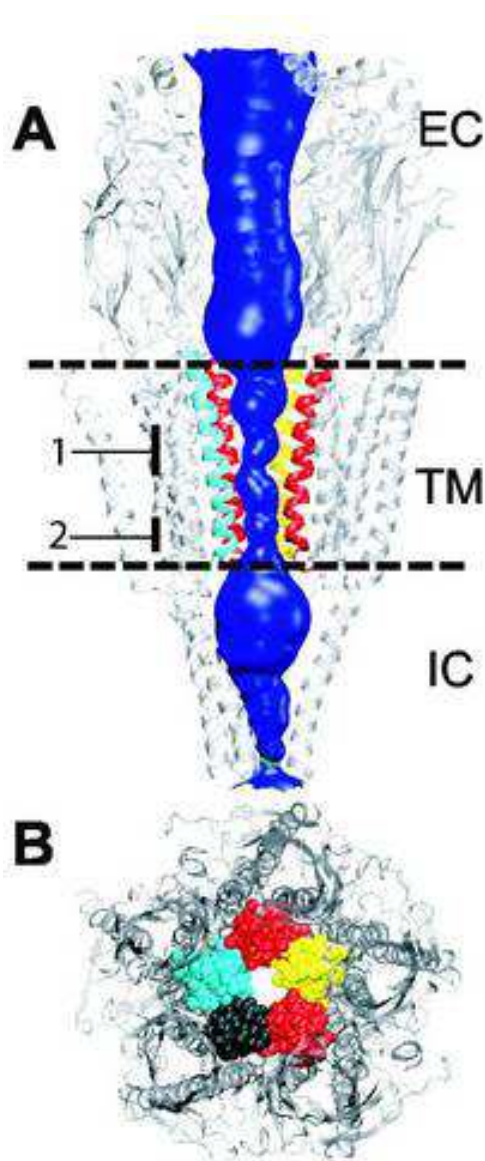}
    \caption{\label{fig:structure}%
      {\textbf{A} Overall architecture of the nAChR (PDB id 2BG9) as determined
      by cryo-electron microscopy \protect\citep{Unwin05}, showing the
      extracellular (EC), transmembrane (TM), and intracellular (IC) domains
      and the surface of the pore (calculated with \textsc{hole}
      \protect\citep{Smart96}). The position of the gate is currently
      intensely debated with some evidence pointing to the region of the
      hydrophobic girdle (\textbf{1}) and other experiments indicating the
      constriction close to the intracellular domain (\textbf{2}).  The
      horizontal lines indicate the approximate location of the lipid bilayer.
      The M2 helices are colored by subunit: $\alpha$ (red), $\gamma$
      (yellow), $\delta$ (cyan); the $\beta$ subunit is omitted for clarity
      but shown in black in the top view below.  \textbf{B} Structure of the
      nAChR TM domain viewed down the pore axis from the extracellular end,
      with the M2 helices in space-filling format (images produced with
      \textsc{vmd} \citep{VMD} and \textsc{Raster3D} \citep{Raster3D}).}
}
\end{figure}

The mechanism of gating of ion channels is a central problem in
membrane protein biophysics. One class of channel for which structural
and biochemical data are available is the one containing the
ligand-gated ion channels
(LGIC\footnote{%
  Abbreviations: LGIC, ligand-gated ion channels; nAChR, nicotinic
  acetylcholine receptor; 5HT$_{3}$R, 5-hydroxytryptamine (serotonin)
  receptor; GABA$_\text{A,C}$R, $\gamma$-aminobutyric acid receptor
  type A or C; GlyR, glycine receptor; ACh, acetylcholine; gA,
  gramicidin A; cryo-EM, cryo-electron microscopy; TM, transmembrane;
  MD, molecular dynamics; PMF, potential of mean force; WHAM, weighted
  histogram analysis method; MscS, mechanosensitive channel of small
  conductance
}%
), represented by the nicotinic acetylcholine receptor (nAChR). This
cation-selective channel is made up from five homologous subunits
($\alpha_{2}\beta\gamma\delta$ in muscle type receptors) packed around
a central pore, forming a structure with fivefold pseudo-symmetry
\citep{Lester04R}. Cryo-electron microscopy studies of the
\textit{Torpedo} nAChR \citep{Miyazawa03,Unwin05} have revealed three
domains: an extracellular domain containing the neurotransmitter
binding site; a transmembrane (TM) domain forming a pore across the
lipid bilayer; and an intracellular domain providing binding sites for
cytoskeletal proteins (figure~\ref{fig:structure}A).  The structure of
the extracellular ligand-binding domain is homologous to that of a
water-soluble acetylcholine binding protein \citep{Brejc01}. Each
subunit of the TM domain contains four membrane-spanning helices, M1
to M4. The five M2 helices come together to form an approximately
symmetrical TM pore (figure~\ref{fig:structure}B). The lower
(intracellular) half of the pore is formed by a polar 3~\AA{}
constriction (lined by serine and threonine residues), whilst the
upper (extracellular) half of the pore contains a more hydrophobic
constriction of the same radius, lined mainly by valine and leucine
residues. The protein was crystallized in the absence of
acetylcholine; thus the receptor is presumed to be in a functionally
closed state. Despite this, the narrowest zone of the pore is
sufficiently wide to accommodate e.g.{} three water molecules or a
Na$^{+}$ ion and two waters side by side. Thus, there is an apparent
paradox of a functionally closed pore which is not fully occluded.
This contrasts with e.g.{} the structure of the bacterial K$^{+}$
channel KirBac1.1 \citep{Kuo03} where the putative gate has a radius
of 0.5~\AA{}, thus sterically occluding the ion passageway.  However,
the situation is already less clear cut for KcsA \citep{Zhou01},
another bacterial K-channel, whose gate constriction is just wide
enough to admit a bare potassium ion. Interestingly, those
constrictions are also formed by hydrophobic residues.

A possible answer to the paradox lies in the concept of
\emph{hydrophobic gating} of ion channels, which has been postulated
in the literature \citep{Cohen92,Unwin95,Chung98,Hille01,Moe00} and
was quantitatively developed on the basis of computer simulations of
simplified pore models \citep{Beckstein01,Beckstein04,Beckstein04a}. A
hydrophobic gate is a constriction that acts as a desolvation barrier
for ions. It is so narrow (pore radius $R < 4$~\AA) that an ion has
to shed at least some water molecules from its hydration shell if it
were to pass the constriction.  Because this requires a large amount
of free energy (the solvation free energy for a potassium ion is about
$-308\,\text{kJ}\,\text{mol}^{-1}$ and for a sodium ion it is
$-391\,\text{kJ}\,\text{mol}^{-1}$ \citep{Schmid00}), passage of the
ion is energetically unfavorable and thus blocked, even though the
geometry would permit permeation of a partially hydrated ion or a
water molecule. The desolvation barrier is only effective if hydration
shell water molecules cannot be temporarily substituted with e.g.{}
hydroxyl groups from side chains or the protein backbone as seen, for
instance, in the narrow ($R \approx 1.5$~\AA) selectivity filter of K
channels \citep{Zhou01,Berneche01}.  Thus, the constriction has to be
lined by hydrophobic side chains, whose methyl groups will not
participate in solvating an ion. %
\begin{figure}[tb]
  \centering
  \includegraphics[clip,width=120mm]{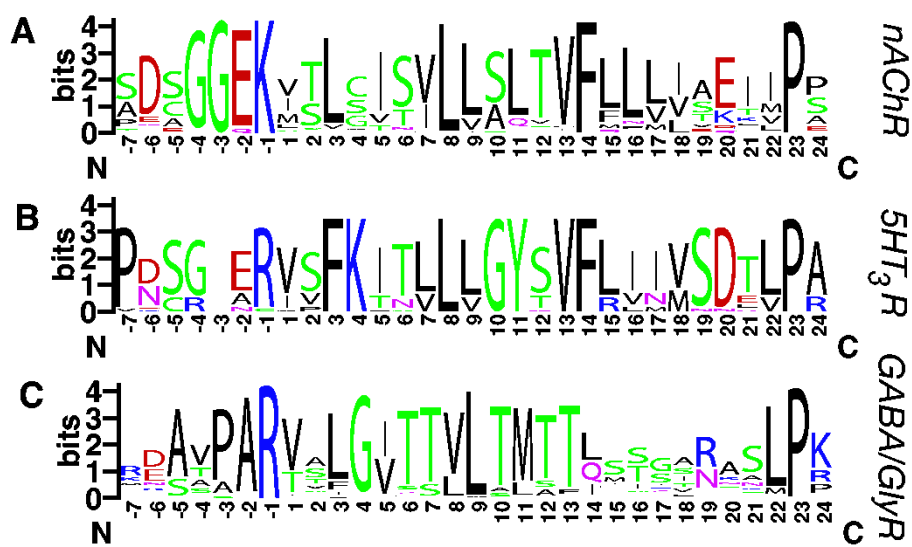}
  \caption{\label{fig:sequencelogos}%
  {Sequence logos\index{sequence logo} \protect\citep{Schneider90} for the
    ligand gated ion channels in the M2 region. \textbf{A} nAChR. \textbf{B}
    5HT$_{3}$ (serotonin) receptors. \textbf{C} GABA and Glycine receptors.
    Acidic residues are marked red, basic ones blue, polar ones green.
    Hydrophobic residues (and proline) are colored black. Residues are
    counted within M2, using the `prime' nomenclature where $\alpha$M243 is
    designated $1'$. %
    The information content per residue is measured in bits; a totally
    conserved residue has an information content of $4.32$~bits. The size of a
    letter indicates this amino acid's contribution to the information content
    at the position.}
}
\end{figure}
According to the electron microscopy structure \citep{Miyazawa03} and
biochemical studies \citep{Bertrand93a,Karlin02}, the central portion
of the nAChR pore is lined by valine and leucine side chains.
Analysis of Cys-loop receptor M2 sequences in terms of sequence logos
\citep{Schneider90}, shown in figure~\ref{fig:sequencelogos}, reveals
that the LxxxVxxxV/L motif is well conserved amongst nAChR receptors
(residues $9'$, $13'$, and $17'$ when numbered from the N-terminal end
of the M2 helix), and is replaced by an equivalent LxxxVxxxI motif in
5HT$_{3}$ receptors \citep{Lester92,Bertrand93a,Galzi94,Labarca95},
hinting at the biological importance of these hydrophobic residues.
There is also evidence that a similar motif is conserved in
prokaryotic homologues of the nAChR \citep{Tasneem05}, suggesting that
this was an early evolutionary feature of this family of ion channels.
Here we argue that these hydrophobic residues form a \emph{hydrophobic
  gate} (called the `hydrophobic girdle' by \citet{Cohen92} and later
Unwin and colleagues \citep{Unwin95,Miyazawa03}).
In GABA and glycine receptors the LxxxVxxxV/L motif is replaced by a
LxxxT motif (residues $9'$ and $13'$), suggesting the nature of the
gate may be modified in anion-selective channels.

However, the position of the gate is highly contentious, with
different experiments pointing towards different locations (see
figure~\ref{fig:structure}).  As discussed below, some studies employing
methods probing the accessibility of pore lining residues find the
gate close to the intracellular end of the pore
\citep{Akabas94,Wilson98,Wilson01,Karlin02,Paas05} whereas the EM
structure \citep{Miyazawa03} and the same accessibility method applied
to 5HT$_{3}$R \citep{Panicker02} point to the hydrophobic
girdle as the gate; other mutation studies in LGIC point to the
importance of the ring of conserved Leu and Val residues, suggesting a
gate closer to the extracellular end of the pore
\citep{Revah91,Filatov95,Labarca95,England99,Horenstein01,Plazas05}. %
Knowing the position of the gate is crucial in the larger enterprise
of understanding the full gating behavior of the ligand gated ion
channels as a paradigm for a complex receptor structure-function
relationship \citep{Changeux98R}.  The simulations presented in this
work are answering the question whether a hydrophobic girdle (as seen in
the EM structure) could act as a barrier to ion permeation and provide
quantitative evidence for the qualitative idea of a hydrophobic gate.

For the bacterial mechanosensitive channel MscS, the concept of
hydrophobic gating has already been investigated using molecular
dynamics computer simulations. \citet{Anishkin04} conclude that the
crystal structure, which exhibits a hydrophobic constriction of radius
4~\AA{} \citep{Bass02}, represents a closed state because water (and
ions) do not enter the putative gate; \citet{Spronk06} challenge this
view because in their simulations both ion and water fill the pore
once an external transmembrane potential is applied across the
channel.  Continuum electrostatics calculations on a simplified model
of KcsA \citep{Chung98} and the crystal structure
\citep{Roux00,Jogini05} indicate that the intracellular hydrophobic
constriction presents a high dielectric barrier even though it does
not sterically occlude the ion pathway. Recent continuum
electrostatics \citep{Corry04a} and short equilibrium MD and Brownian
dynamics calculations \citep{Corry05} suggest that hydrophobic gating
may also apply to nAChR.  However, comparisons with estimates of free
energy profiles for ion permeation through model pores based on
atomistic simulations \citep{Beckstein04a} showed that continuum
approximations can be inaccurate for pores of sub-nanometer
dimensions, especially when hydrophobic effects (which are collective
effects of the solvent) are involved. Thus, such methods cannot
readily provide a rigorous test of the hydrophobic gating hypothesis.
Furthermore, equilibrium simulations can only show what would happen
under equilibrium conditions within the time span of the simulation
(i.e.{} $1$ to $100$~ns, this time scale still being perhaps three
orders of magnitude short of that of channel gating).  For instance,
for an \emph{open} state conductance of $50$~pS as typical for nAChR
\citep{Hille01} one would expect on average about one ion permeation
event per about $60$~ns in equilibrium (using rate theory to estimate
the equilibrium flux: start from the rate theory flux $\Phi = I/q
\approx \Phi_{0} \sinh(qV/2kT)$ \citep{Beckstein04}, expand near
equilibrium, i.e.{} vanishing driving potential $qV \ll kT$, and using
$I=gV$ arrive at a rough estimate for the equilibrium flux $\Phi_{0}
\approx 2kT g q^{-2}
= 0.016\ \text{ns}^{-1}$ or one ion per 63~ns). %
For a \emph{closed} gate the time for such an event to occur would be
at least one order of magnitude larger.  That means that it is
impossible to quantify any sizable energy barriers, such as that in a
closed gate, using straightforward equilibrium MD. In order to probe
such a region, i.e.\ to estimate reliably the height and extent of the
energy barrier to ion permeation in the nAChR pore, and thus to test
the hydrophobic gating hypothesis, detailed atomistic free energy
calculations are required. Although those `potential of mean force'
calculations are computationally expensive there are a number of
comparable simulations that have been used to determine quantitatively
the energy landscape experienced by K$^{+}$ ions within the
selectivity filter of KcsA \citep{Berneche01} and the gramicidin A
(gA) pore \citep{AllenT03,AllenT04,AllenT06}.

We present the results of such calculations for the height of the
energetic barrier associated with the closed gate of the nAChR, and
show that they support the hydrophobic gating hypothesis. Furthermore,
we evaluate the robustness of this conclusion to the forcefield
employed. These results demonstrate the value of a theoretical
physical approach to a mechanistic biological problem that is
difficult to address via direct experimentation.

\section{Methods}
\label{sec:methods}

\begin{figure}[tb]
  \centering
  \includegraphics[clip,width=87mm]{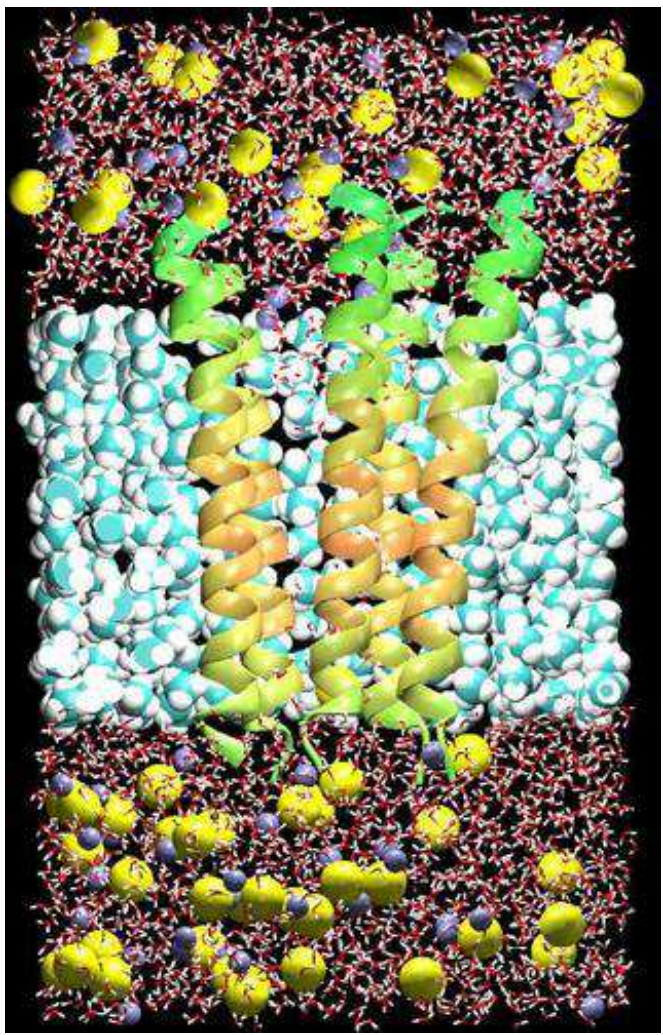}
  \caption{\label{fig:system}%
    {Simulation system. The M2 helix bundle is depicted as ribbons (from the
    cryo-electron microscopy structure of the nAChR TM domain, pdb 1OED
    \protect\citep{Miyazawa03}), the membrane mimetic slab is made from
    methane molecules on a fcc lattice, water molecules are represented as
    sticks, and ions as ice-blue (Na$^{+}$) or yellow (Cl$^{-}$) spheres.}
}
\end{figure}

In order to determine the equilibrium distribution and the potentials of mean
force of water and ions within the pore of the nAChR, we employed fully
atomistic molecular dynamics simulations of the pore-forming M2 helix bundle,
embedded within a membrane-mimetic slab of methane-like pseudo-atoms, with
water and ions (Na$^{+}$ and Cl$^{-}$) equivalent to a concentration of 1.3~M
on either side of the slab (figure~\ref{fig:system}).

\subsection{Model}
\label{sec:model}

The initial model of the M2 bundle consists of residues $\alpha$E241
to $\alpha$V271 (and corresponding residues on the other chains) from
the 4~\AA{} resolution cryo-EM structure (PDB id 1OED
\citep{Miyazawa03}). All titratable residues are in their default
(charged) state at pH 7 (as predicted by p$K_{a}$ calculations on the
whole TM domain using \textsc{WhatIf} \citep{Vriend90,Nielsen01} with
\textsc{DelPhi} \citep{Rocchia01} as the Poisson-Boltzmann solver)
except for the helix termini, which were kept neutral to minimize
their influence on the system.  The M2 bundle is embedded in a
bilayer-mimetic slab of thickness 3.2 nm, made from CH$_{4}$
molecules, which are held on a face-centered cubic lattice (cubic
lattice constant 0.75~nm) by harmonic restraints of strength $k_{0} =
1000\ \text{kJ}\,\text{mol}^{-1}\,\text{nm}^{-2}$. %
The total system size was 15298 atoms, made up of 2819 protein atoms, 2974
water molecules, 689 `membrane' CH$_{4}$ molecules, and 112 ions. %
The pore was initially solvated with water. Cations and anions were
added alternately by exchanging a water molecule at the position of
minimum potential energy for the ion, leading to an approximately
uniform distribution of ions.
For the protein and CH$_{4}$ the OPLS all-atom force field \citep{Kaminski01}
was used but for a sensitivity analysis we also performed simulations with the
united-atom GROMOS96 force field \citep{Gromos96}.  For water the SPC model
was employed \citep{Berweger95}, and ion parameters were taken from the work
of \citet{Aqvist90}. During the simulations, the protein backbone was
harmonically restrained with force constant $k_{0}$ whilst the side chain
atoms, water and ions were free to move, resulting in positional root mean
square fluctuations (average C$\alpha$ fluctuation $0.3$~\AA) similar to 
those seen in simulations of the TM helix bundle in a lipid bilayer
\citep{Hung05} (ca.\ $0.6$~\AA, excluding mobile loops). 

\subsection{Simulation Details}
\label{sec:sim}

Simulations were performed with \textsc{gromacs} 3.2.1 \citep{Lindahl01} at
constant temperature (300~K or 27$^{\circ}$C) and pressure (1~bar) normal to
the membrane, using weak temperature (time constant $\tau = 0.1$~ps) and
pressure coupling ($\tau = 1$~ps) algorithms. Electrostatic interactions were
accounted for by a particle mesh Ewald method \citep{Dar93} (real space cutoff
1~nm, grid spacing 0.15~nm, fourth order interpolation) whereas van der Waals
interactions were computed within a cutoff of 1.4~nm. Protein bonds were
constrained with the LINCS algorithm \citep{Hess97} whereas the SPC water
molecule bonds were constrained with SETTLE \citep{Miyamoto92}. The MD time
step was 2~fs.

The equilibrium simulation in the OPLS-AA force field was performed
for 60~ns; GROMOS96 equilibrium simulations were run for 80~ns.  The
single-channel conductance under physiological conditions of the
(open) \textit{Torpedo} nAChR is 30 to 50~pS \citep{Hille01},
corresponding to an equilibrium mean passage time per ion of about
60~ns. Thus, as mentioned in the Introduction, a 60~ns simulation
should enable us to sample the distribution of ions within an M2
bundle pore if it were permeable to ions (though in any case one would
not expect to observe permeation events).

\paragraph{Potentials of mean force}

Potentials of mean force (PMF) for permeant species were obtained from
a combination of equilibrium and umbrella sampling
\citep{Torrie77,Valleau77} simulations. A starting configuration for
each umbrella window was obtained from the equilibrium simulation by
selecting a frame from the trajectory that had a particle of interest
within about 1~\AA{} of the window center, or by exchanging a suitable
water molecule with an ion. Thus neighboring window configurations
tend not to be correlated.  For umbrella sampling, the particle of
interest was harmonically restrained to subsequent positions on the
channel axis (with typical values of the restraining force constant
$1558\ \text{kJ}\,\text{mol}^{-1}\,\text{nm}^{-2}$) in 101 or more
windows of width $\Delta z=0.396$~\AA{} for Na$^{+}$, Cl$^{-}$, and
water (see table S1 in Supplementary Material for a detailed listing
of umbrella sampling parameters). %
This choice of umbrella sampling parameters allows the particle of
interest to diffuse into neighboring windows (the energy required to
do so is only about $1.5\,kT$, i.e.\ a typical thermal fluctuation),
leading to good overlap between windows \citep{AllenT03}. Each window
simulation was typically run for 1.2~ns, with the initial 0.2~ns being
discarded as equilibration time; hence a single PMF comprises 101~ns
of simulation time or about 50 million configurations of the sampled
particle (at a simulation time step of 2~fs). %
The umbrella sampling simulations for \NA{} with the OPLS force field
were extended into the bulk regions with a harmonic flat-bottomed
cylindrical confinement potential \citep{AllenT04} (radius $1.3$~nm,
force constant $4500\ \text{kJ}\,\text{mol}^{-1}\,\text{nm}^{-2}$). 25
windows were added on the intracellular side ($\Delta z = 0.4$~\AA,
$k=1527\ \text{kJ}\,\text{mol}^{-1}\,\text{nm}^{-2}$) and 20 on the
extracellular side ($\Delta z = 0.5$~\AA, $k=977\ 
\text{kJ}\,\text{mol}^{-1}\,\text{nm}^{-2}$), so that the \NA{} PMF
comprises of 146 windows over a length of 60~\AA.

Resulting histograms were unbiased using the weighted histogram
analysis method \citep{Kumar92} (WHAM), with 300 to 500 bins and a
tolerance of $10^{-5}\,kT$ for the individual window offsets. The PMFs
are converged with respect to the number of bins and the tolerance;
more details can be found in the Supplementary Material.  PMFs were
constructed from matching the umbrella-sampled PMF to the PMF derived
from the equilibrium density (i.e.{} the Boltzmann-sampled PMF)
\begin{equation}
  G(z) = -kT \ln \frac{n(z)}{n_{0}} + C
  \label{eq:equilibriumPMF}
\end{equation}
in the mouth regions of the pore ($n(z)$ denoting the average density
in the pore along the $z$-axis and $n_{0}$ the density in the bulk,
$C$ being an undetermined constant not relevant for our
discussion). Both methods produce overlapping results in the mouth
region, even though combining them is not a rigorously defined
operation (see below).

Many ion channel properties are discussed in terms of the
one-dimensional PMF along a single reaction coordinate (also referred
to as the free energy profile) \citep{Hille01}. However, as
\citet{Roux04} point out, there are a number of theoretical/conceptual
and practical problems associated with the free energy profile. It is
only well defined in the pore region because in the bulk the motion of
the particle is not bounded orthogonal to the reaction coordinate
(although in MD simulations periodic boundary conditions ensure
arteficial confinement). Practically, one can umbrella-sample into
the bulk/mouth regions with the help of a cylindrical confinement
potential. This procedure was employed for the OPLS \NA{} PMF and
resulted in a better converged PMF. On the practical side, one cannot
hope to sample the full multi-ion PMF because all movements of ions
are `slow' degrees of freedom.  With a 1D reaction coordinate
approach only one slow degree of freedom (such as the movement in
$z$-direction) can be controlled; the other slow degrees will not
equilibrate on the time scale of the umbrella sampling simulation and
therefore should be controlled by other means (see \citep{Roux04} for
a full discussion). A well defined quantity is the one-ion PMF, which
is obtained as the PMF of one particular ion while no other ions are
present in the pore. From the one-ion PMF one can, for instance,
compute the maximum single channel conductance, equation~\ref{eq:gmax}. If
the other ions are not excluded from the pore by a computational
device such as an exclusion potential \citep{AllenT03,AllenT06} one
can resort to only include those frames of a trajectory in the WHAM
analysis during which the one-ion condition is fulfilled. The latter
approach was chosen for the OPLS \NA{} PMF. The one-ion region
$\mathcal{P}_{1}$ was defined as $-21.6\ \text{\AA} \leq z \leq 6.4\ 
\text{\AA}$ (roughly corresponding to the region between the $1'$ and
the $17'$ position; see figures~\ref{fig:density} and \ref{fig:pmf}) on
the basis that the ionic equilibrium densities
(figure~\ref{fig:density}) drop sharply inside this region and, that
after analysis of umbrella windows, only 14 out of 72 windows showed
multiple ion occupancies in this region. Those windows were simply
rerun after exchanging the offending ion(s) with a bulk water or after
assigning new initial velocities. Both approaches yielded simulations
with only the sampled ion in $\mathcal{P}_{1}$. (Details can be found
in the Supplementary Material.)

We also calculated a PMF across the whole pore by including the
additional windows outside $\mathcal{P}_{1}$. Although this is not a
rigorous approach it turns out that including the multiple-occupancy
regions in the WHAM procedure does not alter the one-ion part of the
PMF. Because WHAM is a rather sensitive `global' fitting procedure
this gives at least an indication that our PMFs are reasonably robust
quantities, even without the full rigorous treatment. The
Boltzmann-sampled PMF is by definition the multi-ion PMF (though it
might not be converged); our umbrella-sampled PMF contains multi-ion
components in the mouth region and the one-ion PMF in
$\mathcal{P}_{1}$. We join those two PMFs because they tend to overlap
rather well in the mouth regions of the pore and because the
umbrella-sampled PMF for water superimposes almost exactly on the
Boltzmann-sampled one (which can easily be sampled across the whole
pore) as shown in figure~\ref{fig:pmf}.

\subsection{Analysis}
\label{sec:methods-analysis}

\paragraph{Simulations}
Pore radius profiles were calculated with \textsc{hole}
\citep{Smart96} from the cryo-EM structure with a probe radius of
$0.14$~nm.

To address the question of the existence of a vapor-lock mechanism we
analyzed the water occupancy $N_{\text{water}}$ of the hydrophobic
constriction (at $13'$, $-2.6\ \text{\AA} \leq z \leq 1.6\ \text{A}$,
see figures~\ref{fig:density} and \ref{fig:pmf}) from the 60~ns
equilibrium simulation. This region was chosen after preliminary
inspection showed intermittent vapor phases only occurring in this
$4.2$~\AA{} section of the pore, reminiscent of results from our
previous work on hydrophobic model pores
\citep{Beckstein01,Beckstein03}. 

Single-channel conductances were estimated from the PMF $G(z)$ as
\citep{Roux04,AllenT04}
 \begin{equation}
   \label{eq:gmax}
   g_{\text{max}} = \frac{e^{2}}{kT L^{2}}\, 
   \biggl(L^{-1}\int_{\mathcal{P}_{1}}\!d{z}\, D(z)^{-1}
   e^{+G(z)/kT}\biggr)^{-1}\,
   \biggl(L^{-1}\int_{\mathcal{P}_{1}}\!d{z}\, e^{-G(z)/kT}\biggr)^{-1}
 \end{equation}
where the averages are carried out over the pore region
$\mathcal{P}_{1}$ where only one ion occupies the pore.  The diffusion
coefficients in the pore were estimated to be half of the experimental
bulk value \citep{Tieleman01} ($D_{\text{bulk}}(\text{Na}^{+}) = 1.33\ 
\text{nm}^{2}\,\text{ns}^{-1}$, $D_{\text{bulk}}(\text{Cl}^{-}) =
2.03\ \text{nm}^{2}\,\text{ns}^{-1}$ \citep{Hille01}).  This
$g_{\text{max}}$ estimate is strictly true only for one-ion channels
(or one-ion PMFs). In addition to $\mathcal{P}_{1}$ we also define the
barrier region $\mathcal{P}_{b}$, $-23\ \text{\AA} \leq z \leq 10\ 
\text{\AA}$ (E$-2'$ to E$20'$), which is the extent of the barrier in
the \NA{} PMF (see figure~\ref{fig:pmf}). The one-ion condition is not
strictly fulfilled but because of the robustness of the one-ion PMF we
use the PMF over $\mathcal{P}_{B}$ as an approximation for the proper
one-ion PMF and calculate $g_{\text{max}}$ across the full barrier.
(Because $g_{\text{max}}$ depends on $L^{-2}$ we use the exact same
definition for the pore regions for all our simulations in order to
compare the conductance estimates.)

\paragraph{Sequences}
An alignment of the transmembrane domain of the ligand gated ion channels was
prepared, based on the Pfam (`protein family') PF02932
(Neurotransmitter-gated ion-channel transmembrane region), which contains 616
members (version 14.0 of Pfam \citep{Bateman04}).  After removal of
fragmentary sequences, 511 remained, which were aligned with \textsc{ClustalW}
\citep{ClustalW}. A calculation of an average-distance neighbor joining tree
(using \textsc{jalview} \citep{Clamp04}) shows four distinct families (nAChR,
5HT$_{3}$R, GABA$_{\text{A}}$R, GlyR) for the whole alignment. Focusing on the
M2 region three distinct families remain: cation-selective nAChR (234
sequences) and 5HT$_{3}$R (20 sequences), and anion-selective GABA/GlyR (244
sequences).  From the M2 sequences, `sequence logos' \citep{Schneider90}
were created.

\section{Results and Discussion}
\label{sec:results}

\begin{figure}[tbp]
    \centering
    \includegraphics[clip,width=87mm]{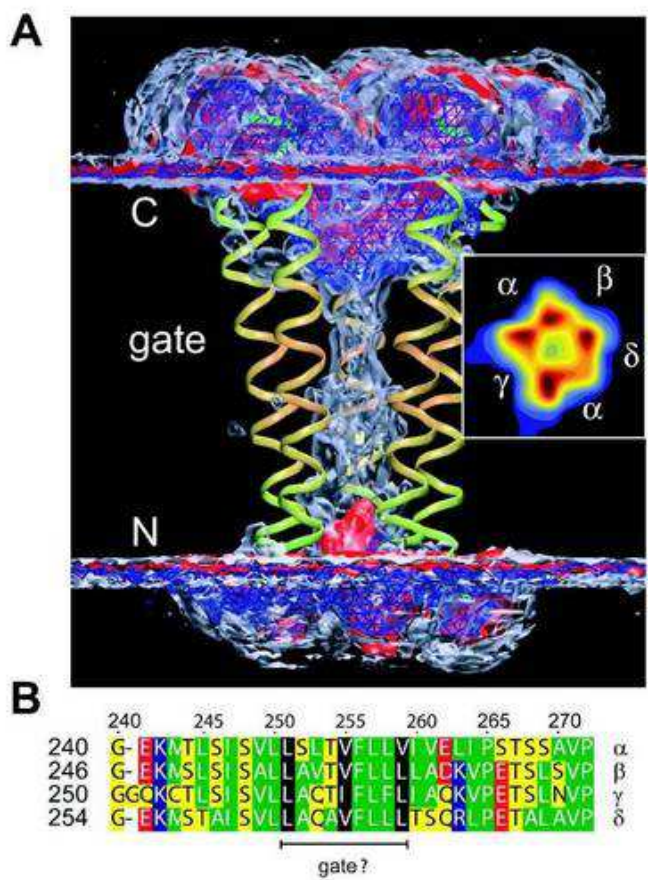}
    \caption{\label{fig:gate}%
      {\textbf{A} Water and spatial ion densities in the nAChR M2
        helix bundle (ribbons; the membrane mimetic slab is omitted
        for clarity). The N and C termini of the helices
        (corresponding to the intracellular and extracellular ends of
        the transbilayer pore respectively) and the hydrophobic gate
        are labeled. The three surfaces show the average density of
        water molecules (gray surface, contoured at 95\% of the bulk
        density), sodium ions (blue surface), and chloride ions (red
        surface), both contoured at 0.1~M. These are taken from a
        60~ns simulation on the M2 helix bundle embedded in a
        membrane-mimetic slab and bathed in a 1.3~M NaCl solution. The
        inset shows a cross-section through the water density in the
        region of the gate. The water density is contoured on a color
        scale, ranging from 5\% (deep blue) to 140\% (deep red) of
        bulk density.  \textbf{B} Sequence alignment for the M2 helix
        sequences from the four subunit types of the \textit{Torpedo}
        nAChR.  The putative hydrophobic gate region, extending from
        L251 to V259 of the subunit, is indicated below the sequence
        alignment.}
}%
\end{figure}

\begin{figure}[tbp]
    \centering
    \includegraphics[clip,width=180mm]{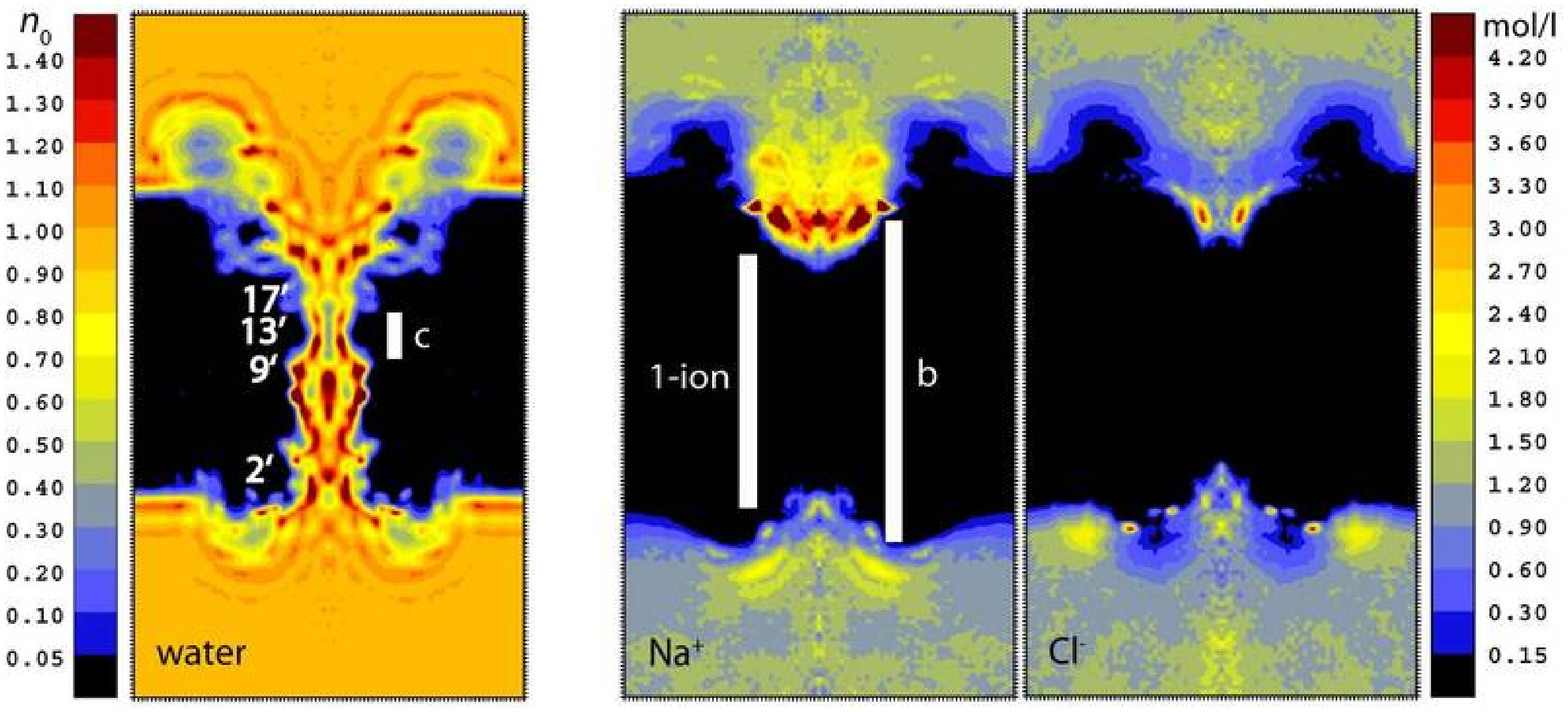}
    \caption{\label{fig:density}%
      {Radially averaged density of water, \NA{}, and \CL{} in the M2
        pore of nAChR (mirrored at the $z$-axis to give an impression
        of the pore environment). Data averaged over one 60~ns
        equilibrium simulation. Water density is given as a fraction
        of the bulk density of SPC water ($n_{0}=0.9669\ 
        \text{g}\,\text{cm}^{-3}$) and the ionic density as a
        concentration. The position of the residues forming the
        hydrophobic girdle ($9'$, $13'$, $17'$) and the $2'$ position (at the
        intracellular constriction) are indicated.  The white bars
        define regions of the pore mentioned in the text (\textbf{c}:
        $13'$ hydrophobic constriction, \textbf{1-ion}: region from
        which a true 1-ion PMF was obtained, \textbf{b}: barrier
        region).}
}%
\end{figure}

The long 60~ns equilibrium simulation reveals that although water
penetrates the full length of the pore, ions do not
(figures~\ref{fig:gate}A and \ref{fig:density}). Instead there is a
local increase (up to $6\times$) in concentration of cations at the
extracellular mouth of the pore.  This reflects the presence of rings
of negatively charged aspartate and glutamate residues at the mouths
of the M2 helix bundle (at the $20'$ and $24'$ position), as can be
seen in figures~\ref{fig:gate}B and \ref{fig:sequencelogos}. By
increasing the local ion concentration near the channel entrance, the
charged rings lower the effective access resistance and so increase
the single-channel current once the pore opens \citep{Hille01}.
 
Ions fail to enter the central hydrophobic section of the M2 pore: at
a bulk concentration of 1.3~M NaCl, the ionic density in the gate
drops to less than $0.1$~M during 60~ns of equilibrium MD. This is
consistent with the hydrophobic gating hypothesis.  However, a 60~ns
duration equilibrium simulation cannot sample the distribution of ions
over an energy barrier of greater than about $5\,kT$ reliably. Thus,
umbrella sampling simulations are needed to estimate the barrier
height and exact position of the residues responsible for the barrier
(see below).

\begin{figure}[tbp]
    \centering
    \includegraphics[clip,width=130mm]{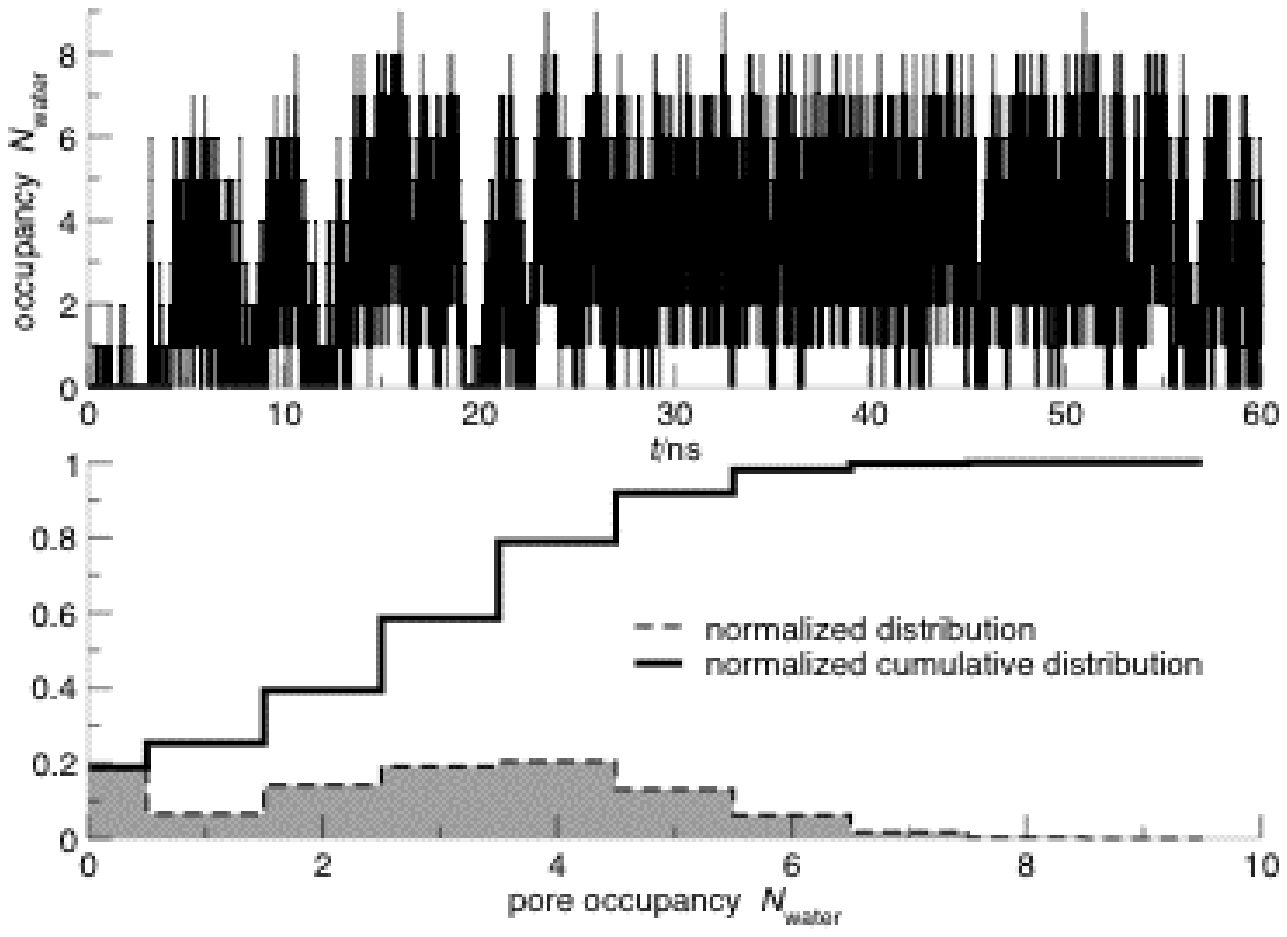}
    \caption{\label{fig:count}%
      {\textbf{Top}: Number of water molecules $N_{\text{water}}$ in
        the hydrophobic constriction at the $13'$ position (pore
        occupancy) over a 60~ns equilibrium simulation.
        \textbf{Bottom}: Normalized distribution of the pore occupancy
        and its cumulative distribution.}
}%
\end{figure}

\subsection{Water in the hydrophobic constriction}
\label{sec:watergate}

The presence of the peptide backbone makes the nAChR pore
significantly more polar than hydrophobic model pores studied in our
previous simulations \citep{Beckstein03,Beckstein04,Beckstein04a}. It
is therefore not surprising that water is seen all along the pore
(figure~\ref{fig:density}). The putative gate region around
$\alpha$V$13'$ (region \textbf{c} in figure~\ref{fig:density}) is of
particular interest, in that the water density is localized in an
approximately pentameric arrangement, forming a hollow tube with on
average little water in the center (inset of figure~\ref{fig:gate}A).
This hints at strong constraints on water positions: water may only
maintain its presence if it interacts in a fairly restricted
configuration.  Inspection of the trajectory reveals that the
pentameric distribution only emerges on averaging. It does not reflect
structures such as highly ordered five-membered rings. On occasions, a
hydrogen-bonded string of water molecules can be seen to sample the
preferred locations though mostly the pore is simply water filled
(also see the supplementary figure S4 and the movie at
\url{http://sbcb.bioch.ox.ac.uk/oliver/download/Movies/watergate.mpg}).
In MD simulations of water in MscS, extended vapor phases with some
intermittent water filling were observed in the putative gate region
\citep{Anishkin04,Sotomayor04,Spronk06}. Partly on this basis,
\citet{Anishkin04} concluded that the MscS structure represents a
closed state and hypothesized that a similar `vapor-lock' mechanism
might be at work in nAChR.  As figure~\ref{fig:count} shows, the
number of water molecules in the constriction fluctuates between zero
and nine over 60~ns. For 75\% of the time, the hydrophobic
constriction is filled with two to nine water molecules and for 19\%
it is void of water (and as such our results are more similar to what
\citet{Sotomayor04} found for water in the MscS hydrophobic
constriction). The distribution in the lower panel of
figure~\ref{fig:count} resembles data obtained for hydrophobic model
pores \citep{Beckstein03}; in particular, it is bi-modal with a
meta-stable vapor-like state ($N_{\text{water}}=0$) and a stable
liquid-like state ($N_{\text{water}}>1$). (In Supplementary Material
we discuss the hydrophobic constriction as a simple hydrophobic nano
pore.) These data indicate that extended vapor phases are not
responsible for blocking ionic current, simply because those phases
are not sufficiently stable. Nevertheless, the appearance of a vapor
state is an indicator of the hydrophobic nature of the pore, which, as
we will demonstrate below, in itself can already be sufficient to
create a desolvation barrier for \NA.

\subsection{Position of the gate}
\label{sec:pmf}

\begin{figure}
    \centering
    \includegraphics[clip,width=100mm]{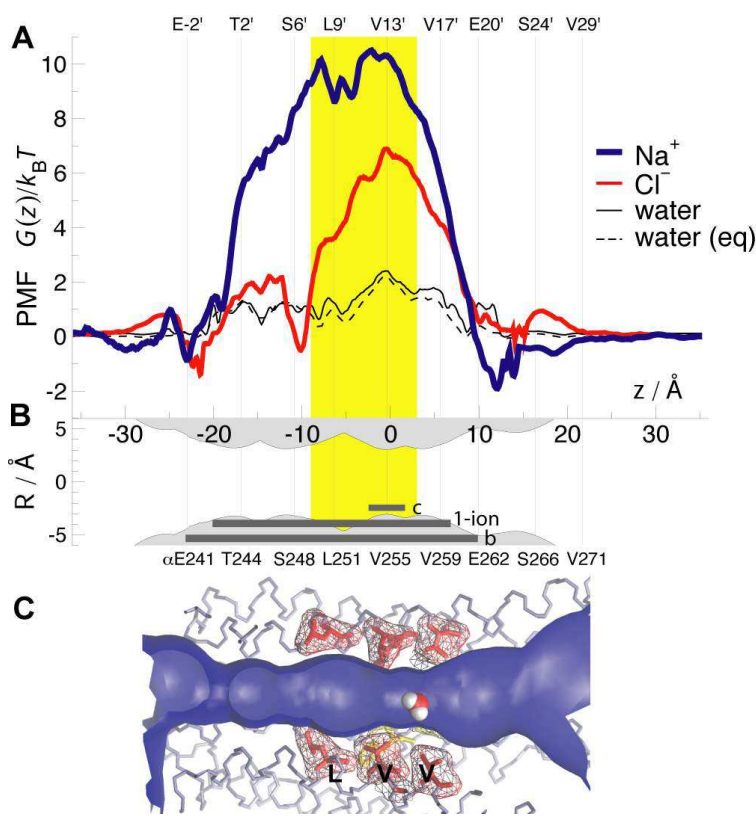}
    \caption{\label{fig:pmf}%
      {Free energy and radius profiles of the M2 bundle pore.
        \textbf{A} Potentials of mean force (PMFs) for water (black
        line), Cl$^{-}$ (red line), and Na$^{+}$ (blue line) along the
        $z$ axis of the M2 pore (from umbrella-sampled simulations).
        The dashed line is the PMF for a water molecule in the pore,
        derived from the density of one 60~ns equilibrium simulation
        (equation~\protect\ref{eq:equilibriumPMF}). The intracellular
        end of the M2 pore is at $z\approx -25$~\AA; the extracellular
        end is at $z \approx +15$~\AA. The putative gate region runs
        from $z \approx -9$~\AA{} to $+5$~\AA{} (marked by the yellow
        band).  \textbf{B} The \textsc{hole} pore radius profile along
        the $z$ axis.  The sequence numbering at the top of \textbf{A}
        is using the `prime' nomenclature commonly applied to M2; the
        equivalent residue numbers in the \textit{Torpedo} subunit
        sequence are given below \textbf{B}. Gray bars indicate the
        same pore regions as in figure~\protect\ref{fig:density} (c:
        hydrophobic constriction, 1-ion: region with true 1-ion PMF,
        b: barrier region).  \textbf{C} The pore lining surface,
        scaled and aligned so it corresponds to the pore profiles in
        \textbf{A} and \textbf{B}.  The side chains of residues
        corresponding to L251, V255, and V259 of the $\alpha$ subunit
        are shown, as is a water molecule (for purposes of comparison
        only) in space-filling format within the pore.}}
\end{figure}

From the water equilibrium distribution it is possible to estimate
(via a Boltzmann transformation, equation~\ref{eq:equilibriumPMF}) a
free energy profile (i.e.{} PMF) for water along the pore axis. This
direct estimate coincides exceptionally well with the PMF obtained by
umbrella sampling (see figure~\ref{fig:pmf}A). Water molecules do not
encounter significant barriers (the highest one is $2\,kT$ at the
central residue $\alpha$V255 of the hydrophobic girdle), again
indicating that the closed-state nAChR pore is largely water filled. %
Because a vapor-lock mechanism of gating does not seem to be an
accurate description, a more detailed investigation of the energetics
of ion permeation such as the full PMFs of the ions is required.

The same umbrella sampling procedure can be used to obtain PMFs for
ions (both Na$^{+}$ and Cl$^{-}$) along the pore axis
(figure~\ref{fig:pmf}A).  Na$^{+}$ ions encounter a significant
barrier of about $10.5\,kT$. An estimate of the single-channel
conductance for Na$^{+}$, based on the PMF (equation~\ref{eq:gmax}),
yields $g_{\text{max}}=0.33$~pS (table~\ref{tab:gmax}). This is much
lower than the experimentally observed open state conductance of
$30$~pS to $50$~pS \citep{Hille01}. In fact, the conductance for the
whole channel would be even smaller as our estimate does not include
the resistance encountered in other regions of the receptor and the
access resistance at the mouths \citep{Chi99}. As our computed
$g_{\text{gmax}}$ for sodium, which is an upper bound on the true
single channel conductance, is already much smaller than the
(experimental) open state conductance we conclude that the M2 pore
appears impermeable to Na$^{+}$ and hence represents a closed
conformation.
The barrier for sodium is wide, ranging over the complete length of
the hydrophobic girdle, which we identify with the gate. Thus, the
hydrophobic gate may be thought of as being \emph{distributed} from
the $\alpha$L251 to the $\alpha$V259 side chain rings.  The peak of
the barrier coincides with a very narrow (radius 3~\AA---see
figure~\ref{fig:pmf}C) and hydrophobic ($\alpha$V255) region of the
pore but the width of the barrier is mainly due to $\alpha$L251, which
lines the central part of the nAChR pore. 

\begin{table}[btp]
  \centering
 \caption{Maximum single channel conductance
   estimates, calculated with equation~\protect\ref{eq:gmax} over the
   region indicated in the table. Throughout the text the value across
   the whole \emph{barrier} region $\mathcal{P}_{B}$ is quoted but the
   \emph{1-ion} region $\mathcal{P}_{1}$
   value is also given for comparison. $k_{\text{bb}}$ denotes the strength
   of the backbone restraints on the backbone of the M2 helices;
   $k_{0}=1000\ \text{kJ}\,\text{mol}^{-1}\,\text{nm}^{-2}$.} 
  \label{tab:gmax}
  \begin{tabular}{llccc}
    \toprule
    force field & $k_{\text{bb}}/k_{0}$ & ion 
        & \multicolumn{2}{c}{$g_{\text{max}}/\text{pS}$}\\
             &  & & barrier & 1-ion\\
    \midrule
    OPLS-AA & $1.0$ 
         & \NA & 0.33 & 1.0\\
      &  & \CL & 21   & 37 \\
    GROMOS96  & $1.0$
         & \NA & 18   & 108 \\
      &  & \CL & 11   & 25  \\
    GROMOS96  & $0.2$
         & \NA & 16   & 87 \\
    \bottomrule         
  \end{tabular}
\end{table}

A previous study computed Poisson-Boltzmann continuum electrostatic
solvation free energy profiles across the nAChR gate region (using the
full transmembrane domain) in order to assess the influence of
changing pore geometry on ion transport properties \citep{Hung05}. For
the cryo-EM like configuration the results differ somewhat from the
fully atomistic, explicit solvent PMF (figure~\ref{fig:pmf}A).  Notably,
the solvation free energy profile shows three peaks at S248, L251, and
V255, with the V255 peak being the smallest. This discrepancy is not
surprising and rather points to the inherent (and well-known
\citep{Corry00,Roux04}) limitations of the continuum approach,
namely a very strong dependence on small pore radii (which will
dominate the result if computed from a single structure), and the
omission of hydrophobic destabilization of the solvent in apolar
regions \citep{Beckstein04}.

Chloride ions encounter a less pronounced barrier of about $6.5\,kT$,
which is also peaked at the hydrophobic girdle.  In the case of the
Cl$^{-}$ PMF shown in figure~\ref{fig:pmf}A the upper-bound estimate
on $g_{\text{max}}$ is rather large with a value of $21$~pS
(table~\ref{tab:gmax}) because the barriers are not very high and not
very wide. According to the PMF, the central cavity (formed mainly by
L$9'$) can stabilize a solvated Cl$^{-}$ ion but not a solvated
Na$^{+}$. It is somewhat surprising that the negatively charged rings
($20'$ and $24'$, called the extracellular ring \citep{Imoto88}) do
not contribute to any appreciable barrier for Cl$^{-}$ even though one
might expect strong repulsion between anions and the negatively
charged glutamates or aspartates. The simulations show that all the
negative charge ($-6e$) is effectively screened by \NA{} ions
(figure~\ref{fig:density}) so that, on average, the ions form a double
layer protruding into the mouth region, with a cloud of cations
enveloping the anions as shown by the 3D density in
figure~\ref{fig:gate}A. This picture is consistent with the fact that
charge selectivity is not conferred by the extracellular ring but
rather by the so called intermediate ring at $-2'$
\citep{Imoto88,Galzi92,Corringer00R,Gunthorpe01}. The PMF, however,
does not exhibit a barrier near $-2'$. This is due to the fact that we
truncated our model at $-2'$ and that the whole intracellular domain of
nAChR is missing from our model. Selectivity requires charged groups
protruding into a confined environment \citep{Wang92} but the $-2'$
residues are exposed to the bulk in our simulations so that their
effect is effectively screened. It should also be noted that the \CL{}
PMF was constructed from umbrella windows that included multi-ion
configurations in the pore, and hence does not constitute a true
one-ion PMF. Judging from the analysis of the \NA{} PMF the difference
is unlikely to be more than a moderate increase in barrier height by
$1$ to $2\,kT$.

\citet{AllenT06} suggest a number of corrections to fully atomistic
PMFs, which include the effect of polarization of the surrounding
lipid hydrocarbon chains and a correction for the finite (but
periodic) system size. In the case of gA, those corrections lowered
the PMF by a few $kT$ \citep{AllenT04,AllenT06}. An additional source
of uncertainty is the absence of the outer helices (M1, M3, M4) and
the intra- and extracellular domain from our model. For a model of the
$\alpha7$ nAChR we showed that the barriers in a continuum
electrostatic free energy profile almost doubled once the ligand
binding domain and the outer TM helices were added to the M2 bundle
\citep{Amiri05}, essentially due to an expansion of the low dielectric
environment experienced by the ion. Because we are primarily
interested in pinpointing the gate, the absolute value of the PMF
(which is needed for accurate $g_{\text{max}}$ estimates) is less
important than the shape and relative heights. The latter will be less
affected by uniformly applied corrections such as the ones discussed
above so that we are still able to relate the peaks in the PMF to the
residues responsible for the barrier. (It is still necessary to
compute the PMF with all-atom simulations with explicit solvent and
correct long range electrostatics as this is the only method that
accounts for all effects (except for polarizability) that are
important in confined geometries \citep{Beckstein04a,Roux04}.)

\subsection{Sensitivity analysis}

\begin{figure}
  \centering
  \includegraphics[clip,width=75mm]{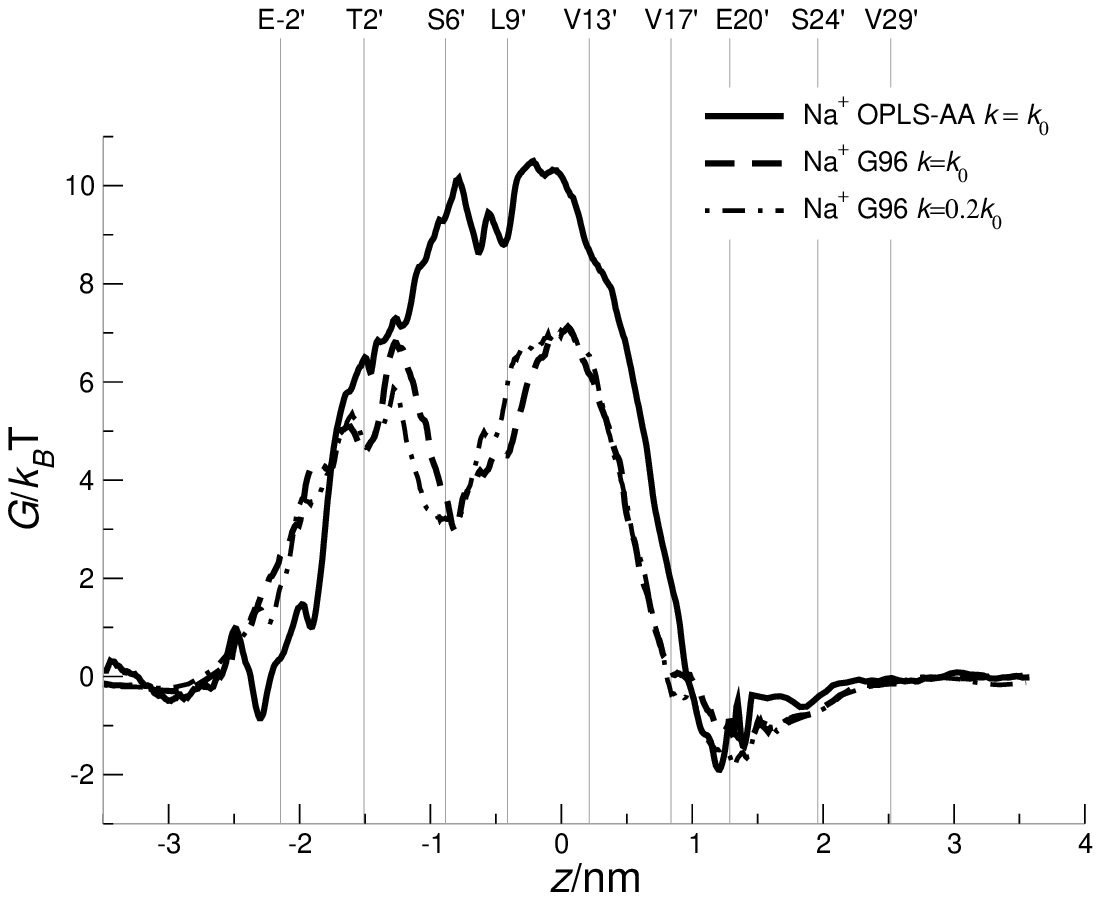}%
  \includegraphics[clip,width=75mm]{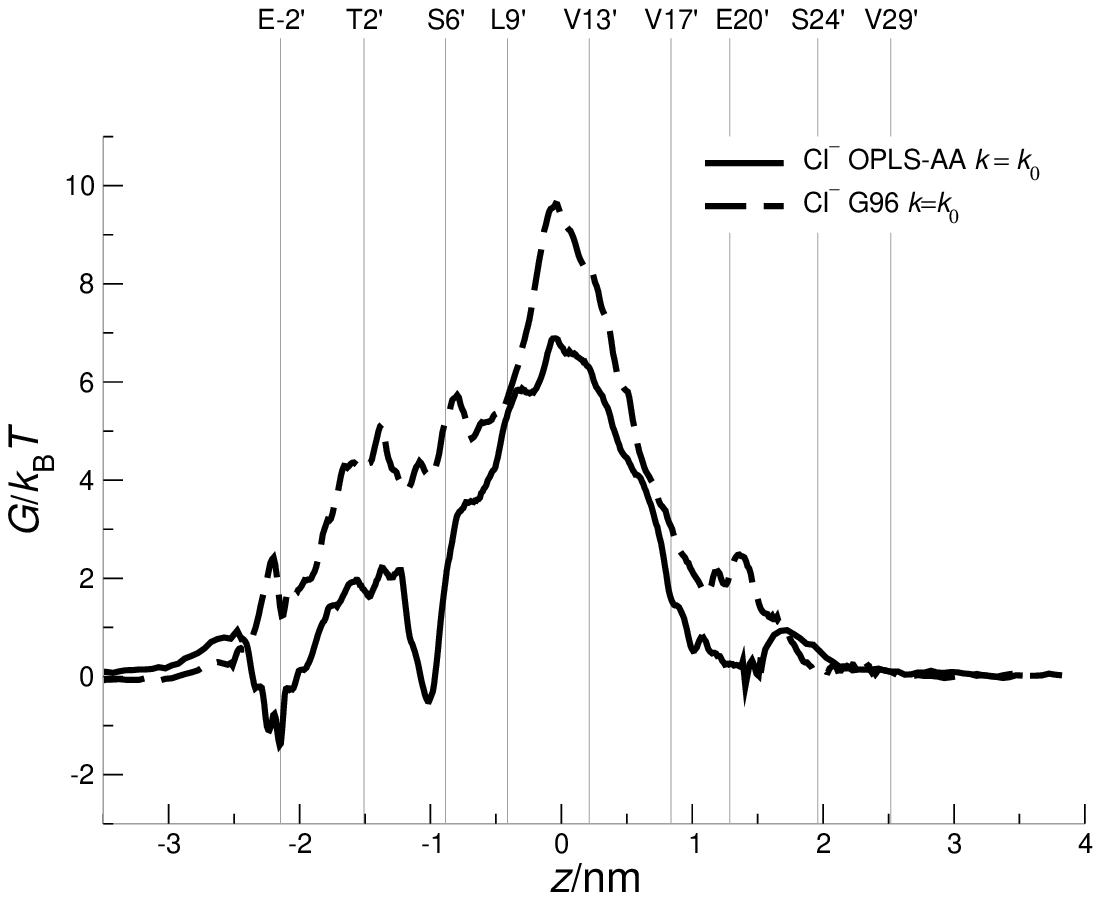}
  \caption{\label{fig:pmf-sensitivity}%
{Sensitivity analysis of the PMF. \textbf{Left}: The PMF for Na$^{+}$
    for the OPLS-AA and the GROMOS96 (G96) force fields (the latter for
    differing strengths of the backbone restraints force constant $k$; see
    text).  \textbf{Right}: The PMF for Cl$^{-}$ ions. In all cases the
    umbrella-sampled and WHAM-unbiased PMF was matched up with the PMF derived
    from equilibrium simulations.}}
\end{figure}

We have tested the sensitivity of the fundamental result, namely that
the L251--V255--V259 side chain rings produce a hydrophobic gate in
the closed state of the nAChR channel, to a number of factors. The
profiles presented in figure~\ref{fig:pmf}A were calculated using the
OPLS all-atom forcefield \citep{Kaminski01}. We have repeated the
umbrella sampling simulations with the united-atom GROMOS96 forcefield
\citep{Gromos96} (figure~\ref{fig:pmf-sensitivity}): The PMFs share
the same qualitative features such as the peak of the barrier at
$\alpha$V255 but the absolute values shift. The \NA{} peak height
reduces to ca.\ $7\,kT$ and the barrier at the lower constriction site
becomes more pronounced.  The united-atom Cl$^{-}$ PMF is higher than
the all-atom one (ca.\ $10\,kT$).  Consequently, the $g_{\text{max}}$
estimates (equation~\ref{eq:gmax}) differ substantially as shown in
table~\ref{tab:gmax}. The sensitivity of PMFs to the force field has
also been noted in the case of gA \citep{AllenT03,AllenT06}.

Our calculations are based on a medium-resolution cryo-EM derived
model \citep{Miyazawa03}.  Thus the coordinates of the M2 helix bundle
are not known with the same precision as from e.g.{} an X-ray
structure.  However, based on the simulations by \citet{Hung05},
relaxing the TM2 structure results in a pore with a radius
$>2.5$~\AA{} so even though the resolution is only $4$~\AA, the pore
(of radius ca.~$3$~\AA) is most likely not an artefact of the lower
resolution and it is always wide enough to admit an ion without
physical occlusion. In the cryo-EM model, pore lining residues only
interact with the solvent in the pore and their M2 neighbours (as the
M2 bundle sits rather loosely in the outer M1/M3/M4 scaffold). Thus,
from the perspective of our simulations, restraining the backbone
atoms should be sufficient to enable sidechains to relax into their
most favourable positions.

To assess the influence of the backbone restraints we also calculated
the Na$^{+}$ PMF based on simulations with a $5\times$ weaker
restraint on the M2 backbone atoms, thus allowing for an enhanced
degree of protein flexibility (GROMOS96 forcefield). The root mean
square deviation (RMSD) of the backbone atoms, averaged over 80~ns of
simulation, is $0.653\pm0.029$~\AA{} for the weaker restraints
compared to $0.388\pm0.014$~\AA{} for the standard restraints. The
RMSD of the side chain atoms is $2.16\pm0.07$~\AA{} (weak restraints)
and $2.48\pm0.11$~\AA, which is of the same magnitude as the side
chain RMSD between the weak and standard restraint simulations
($2.05$~\AA), i.e.{} in both simulations the sidechains are free to
explore a similar range of conformations.  We observed no significant
change in the shape of the PMF (figure~\ref{fig:pmf-sensitivity}) and
the conductances are reasonably similar (table~\ref{tab:gmax}).

We have also investigated the robustness of our results to small
changes in the structural model of the nAChR pore domain. Thus, 
previous preliminary calculations of a Na$^{+}$ ion PMF for the M2
bundle of a homology model of the chick $\alpha$7 nAChR
\citep{Amiri05} yielded a barrier of height between $8$ and $12\,kT$
distributed along the entire TM pore region with
$g_{\text{max}}=0.02$~pS.

Thus the qualitative features of the Na$^{+}$ PMF are robust to small
changes in sequence and in atomic coordinates. This is an important
consideration given the resolution upon which the calculations are
based.  The Cl$^{-}$ PMF differs more noticeably between the two force
fields, which might indicate some subtle difference in how solvation
effects in confined geometries are treated differently by an all-atom
force field (OPLS-AA) versus an united-atom one (GROMOS96).
\citet{AllenT03} compared an all-atom force field (CHARMM PARAM27) and
a united-atom force field (GROMACS) for PMF calculations of the gA
channel. This study (in conjunction with subsequent ones
\citep{AllenT04,AllenT06}) seems to indicate that all-atom force
fields are more successful at describing ion permeation in gA.
Apparently, the same conclusion also holds for nAChR.

\subsection{Comparison with experimental data}
\label{sec:controversy}

All the preceding results are based on the assumption that the
structure of the transmembrane domain of nAChR \citep{Miyazawa03}
faithfully represents the closed state. In fact, nAChR and the other
LGIC do not only have one closed (resting) state but also a closed
desensitized state \citep{Katz57} and Karlin and coworkers find that
the gate is in different positions in these two closed states
\citep{Wilson98,Wilson01}. The cryo-EM images were taken in the
absence of ACh \citep{Miyazawa03} so the EM structure is likely to
represent the closed resting state as desensitization requires
prolonged exposure to agonist; such exposure to the agonist analogue
carbamylcholine resulted in low resolution structures different from
the presumed resting-closed state \citep{Unwin88}.  All EM images were
taken from tubular 2D crystals, which retain the physiological
environment of the receptor: lipid composition and packing of
receptors are close to the situation in the postsynaptic membrane of
the muscle-derived electric organ of the \textit{Torpedo} ray and the
curvature of the tubular crystals is similar to the one of the folds
in the neuromuscular junction \citep{Unw93,Unw95,Unw00,Miyazawa03}. In
addition, the channel in those 2D crystals could be rapidly converted
to the open state through the application of the agonist ACh
\citep{Unw95}, which requires that the initial state was the
resting-closed state \citep{Changeux98R}. Hence it appears unlikely
that the cryo-EM structure represents a desensitized or other
un-physiological state of the receptor. The resolution of the
structure is not very high at $4$~\AA---just good enough to trace the
backbone with some confidence.  But assuming that the backbone is
correctly described, the MD simulations allow the side chains to
re-arrange themselves. Because the PMF is not sensitive to the
strength of the backbone restraints it follows that the local pore
environment, which is formed by the mobile side chains, assumes a
conformation independent of the details of the backbone motions.

The \NA{} PMF (figure~\ref{fig:pmf}) shows a pronounced barrier of about
$10.5\,kT$ at the hydrophobic girdle (between $\alpha\text{L251} = 9'$
and $\alpha\text{V255} = 13'$) and results in a very small maximum
single-channel conductance.  Therefore, the hydrophobic girdle is
identified with the hydrophobic gate of nAChR, as hypothesized by
Unwin and colleagues \citep{Miyazawa03,Unwin03R}, who based their
inference on the protein structure.

This finding disagrees with the results of Karlin and coworkers
\citep{Akabas94,Wilson98,Wilson01,Karlin02} who predict the position
of the (resting) gate at about $\alpha\text{T244} = 2'$, based on
biochemical data (accessibility of substituted cysteines to small,
positively charged, sulfhydryl-specific reagents such as
methanethiosulphonates). However, using the same method in the closely
related 5HT$_{3\text{A}}$ receptor, \citet{Panicker02} find evidence
for a gate between residues $9'$ and $13'$, and further experiments
indicate that the narrow constriction site near $2'$ remains unchanged
between the open and the closed state \citep{Panicker04}---the very
same region that Karlin et al.\
\citep{Akabas94,Wilson98,Wilson01,Karlin02} identify as the gate.
Thus, either the gate location varies between different members of the
same superfamily of ligand gated ion channels, or the methodology is
sensitive to other changes in receptor properties. \citet{Paas05},
using Zn$^{2+}$-binding to engineered His metal binding sites, report
evidence that they interpret as ruling out a hydrophobic gating
mechanism.

The free energy profiles for \NA{} and \CL{} do not allow us to
explain the results of experiments finding the gate near or below the
$2'$ position as we cannot simply extrapolate from monovalent ions to
the reactive reagents such as Zn$^{2+}$ or the methanethiosulphonates.
Further simulations would be required to investigate how those
reagents could penetrate the hydrophobic girdle in the closed state
while being excluded by the narrow constriction near $2'$. In addition
one would need to assess the influence of the mutations to Cys or His
on the channel behavior. The major hurdle in computationally
addressing these experiments in a similar manner as presented here is
the design of force field parameters describing the reactive reagents
in a satisfactory manner.

\section{Conclusion and outlook}
\label{sec:conclusions}

In summary, we have shown that the hydrophobic girdle at the center of
the nAChR M2 pore (as given by the cryo-electron microscopy structure)
acts as a hydrophobic gate. This is in line with recent equilibrium
simulations by \citet{Corry05} and our own previous preliminary
results for the related $\alpha7$ receptor \citep{Amiri05}. However,
it should be noted that the current work rather robustly quantifies
the \emph{free energy} barriers to ion permeation (i.e.{} it includes
explicit entropic and solvation effects in addition to direct particle
interactions) and assesses the influence of choices in the
computational method, namely the force field and the setup of the M2
helix bundle. In trying to evaluate the hydrophobic gating hypothesis,
it is important that a quantitative and robust procedure be employed.
In particular, the accuracy of our calculations is limited only by
accuracy of the force field (not by e.g.{} modeling the water inside
the pore after water in the bulk state, which we know to be a poor
approximation \citep{Hummer01,Beckstein03}) and the structure of the
channel. The cryo-EM structure \citep{Miyazawa03} is the most accurate
atomistic description of a LGIC that is presently available. Our
results demonstrate that if the nicotinic receptor adopts a
conformation as seen in the 4~\AA{} structure then it will (1) block
the flow of \NA{} ions, and (2) it will gate them at the hydrophobic
girdle, and not at the lower constriction site.

The opening of the hydrophobic gate is the final step in the full
gating transition, which involves binding of two ligands,
communicating a conformational change to the transmembrane domain, and
movement of the M2 helices \citep{Grosman00,Miyazawa03,Hung05} as to
increase the pore radius from about $3$~\AA{} to probably not more
than $6.5$~\AA, a radius sufficient to allow ion permeation through a
hydrophobic pore \citep{Beckstein04}.  Hydrophobic gates may be
widespread in a number of ion channels, including K-channels
\citep{Roux00} and bacterial mechanosensitive channels
\citep{Moe00}, in addition to other members of the Cys-loop
superfamily \citep{Lester04R}. A hydrophobic girdle also appears to be
present in the protein-conducting channel formed by SecY
\citep{vandenBerg04} and so may be a general feature of transmembrane
pore proteins that are tightly gated.

Numerous studies have shown that mutations of the conserved leucine
($9'$, $13'$) and valine residues ($13'$, $17'$) to more polar
residues such as serine or threonine affect gating
\citep{Revah91,Filatov95,Labarca95,Lester04R,Plazas05}. The effect of
these mutations tends to be an increase in the opening probability or
in the mean open time, i.e.\ they affect the gating transition.  Hence
the effect of those mutations is more complicated than a simple
increase of conductance of the closed state (an increase in the
leakage current). A possible hypothesis is that the open state is
stabilized by the presence of solvent; protein structure is not
independent of its environment, a fact well known in protein folding.
In this picture, a gate made more polar by a mutation will more
readily admit water and ions, biasing the receptor towards the open
state and so easing the transition to the open state.

From a broader methodological perspective, our results demonstrate the
utility of computational approaches to the interpretation of membrane
protein structures. But they also point to the difficulties obtaining
quantitative output from those simulations.  In the present work the
total simulated time was about $1\,\mu\text{s}$ (more than one order
of magnitude larger than what is currently routinely reported) but as
the analysis with respect to different MD force fields showed, there
is still some work to be done in order to use MD simulations as a
quantitative bridge between structure and function (as seen in
experiments).

\ack Our thanks to all of our colleagues, in particular Nigel Unwin
and Tom Woolf, for valuable discussions. This work was funded by the
Wellcome Trust; Merton College, Oxford; and the EPSRC (via the
Bionanotechnology IRC).

\paragraph{Glossary}
\begin{description}
\item[Free energy] A thermodynamic quantity of a system, which is
  minimal when the system is in equilibrium. It is crucial for a
  quantitative understanding of any physical system.
\item[Gating, Selectivity, Conductance] The three characteristics of
  ion channels \citep{Hille01}. Gating refers to controlling the flow
  of ions through the channel; selectivity measures by how much one
  ion is more likely to permeate than an other; conductance describes
  the ease of flow of ions through the channel.
\item[Ligand gated ion channels, LGIC] A class of ion channels that
  are activated by small molecular ligands such as acetylcholine or
  serotonin. They are primarily involved in functions of the nervous
  system. The gamma-aminobutyric acid type A (GABA$_{\text{A}}$),
  nicotinic acetylcholine (ACh), glycine, and the serotonin
  (5-hydroxytryptamine, 5HT$_{3}$) receptors form a superfamily with
  a common pentameric architecture \citep{Lester04R}.
\item[Molecular dynamics simulation] Atoms are modeled as classical
  particles, moving according to Newton's equations of motions
  $\vec{F}_{i}=m_{i} \frac{d^{2}\vec{x}_{i}}{dt^{2}}$, which are
  integrated numerically with a small time step of typically
  $2\times10^{-15}$~s. The forces are derived as
  $\vec{F}_{i}=-\frac{dU}{d\vec{x}_{i}}$ from a classical force field
  $U$ \citep{FrenkelSmit02}. The output of a simulation is an
  atomically-detailed  `movie'  of the system.
\item[Potential of mean force, PMF] A concept introduced by Kirkwood
  \citep{Kirkwood35}: the free energy of a system as a function of a
  (externally constrained) reaction coordinate \citep{Roux04}. If the
  reaction coordinate is one dimensional (such as $z$ along a channel
  pore) the PMF is called the `free energy profile'. Derived from the
  constrained (configurational) partition function $Z(\xi) \propto
  \int d^{3N}x\, \exp[-U(x)/kT]\,\delta\big(\xi-\hat{\xi}(x)\big)$,
  the PMF along $\xi$ is $\mathcal{W}(\xi) = -kT \ln Z(\xi)$
  \citep{Chaikin00}.
\end{description}

%
%
\newcommand{\newblock}{\relax}


%

\end{document}


%
\title[\runningtitle]{\thetitle}
\author{\theauthor}
\address{\JHU}\ead{orbeckst@jhmi.edu}
\address{\SBCB}\ead{mark.sansom@bioch.ox.ac.uk}
%

\begin{abstract}
  \noindent%
  We present additional data that clarify our choice of the
  `one-ion' region to compute the \NA{} PMF and list some additional
  details of the WHAM unbiasing procedure. We also illustrate the
  behavior of water in the hydrophobic constriction at the $13'$
  position and relate it to our previous work on hydrophobic nano
  pores [Beckstein and Sansom, \emph{Phys.{} Biol.} \textbf{1} (2004),
  42].
\end{abstract}



\submitto{Physical Biology}
\maketitle

\section{Potential of mean force}

The one-ion region was defined after analysis of the $z$-coordinate
trajectories of all ions from the first 101 windows of the \NA{}
umbrella sampling simulations (using the OPLS-AA force field). Six
examples are shown in figure~\ref{fig:UMBzcoord}.

\begin{table}[!b]
  \centering
  \caption[Umbrella sampling parameters]{Parameters of the Umbrella sampling
    simulations. $k_{\text{bb}}$ is the spring constant of the backbone
    restraints on the M2 model of the nAChR TM pore. Umbrella sampling
    of an ion or a water molecule proceeded for a time $T_{\text{1}}$ after an
    equilibration phase of length $T_{0}$. Altogether $N$ windows of length
    $\Delta{z}$ were sampled with a harmonic potential of force constant
    $k$ in the region confined by the $z$ coordinates in the last
    columns. $^{\dagger}$The OPLS-AA \NA{} simulations were carried out with a
    harmonic flat bottomed confinement potential for the ion
    \citep{AllenT04} with radius $1.3$~nm and force constant
    $4.5\,k_{0}$. ($k_{0}=1.0\times10^{3}\  
    \text{kJ}\,\text{mol}^{-1}\,\text{nm}^{-2}$)}
  \label{tab:umbparams}  
  \begin{tabular}{l.l..R..>{$}r<{\leq$}@{$z$}>{$\leq}l<{$}}
    \toprule
    force field & \mmcol{k_{\text{bb}}/k_{0}} & species 
                & \mmcol{T_{1}/\text{ns}} & \mmcol{T_{0}/\text{ns}}
                & \mmcol{N} & \mmcol{\Delta{z}/\text{nm}} 
                & \mmcol{k/k_{0}} & \multicolumn{2}{c}{range$/$nm}\\

    \midrule
     OPLS-AA      & 1.0 & water  &  1.0 & 0.2 & 101 & 0.0396 & 1.558 
                  & -2.5  & 1.5 \\
                  &     & \NA$^{\dagger}$  &  1.0 & 0.2 & 101 & 0.0396 & 1.558 
                  & -2.5  & 1.5 \\
                  &     &        &      &     &  25 & 0.04   & 1.527 
                  & -3.5  &-2.5 \\
                  &     &        &      &     &  20 & 0.05   & 0.977 
                  &  1.5  & 2.5 \\
                  &     & \CL    &  1.0 & 0.2 & 101 & 0.0396 & 1.558 
                  & -2.5  & 1.5 \\
     GROMOS96     & 0.2 & \NA    &  0.5 & 0.1 & 101 & 0.0495 & 0.997  
                  & -2.5  & 2.5 \\
                  & 1.0   & \NA    &  1.0 & 2.0 & 200 & 0.025  & 3.910
                  & -2.5  & 2.5 \\
                  &     & \CL    &  1.0 & 2.0 & 101 & 0.0495 & 0.997
                  & -2.5  & 2.5 \\
    \bottomrule                        
  \end{tabular}
\end{table}

\setlength{\xw}{0.45\linewidth}
\begin{figure}[p]
  \centering\noindent
  \includegraphics[width=\xw]{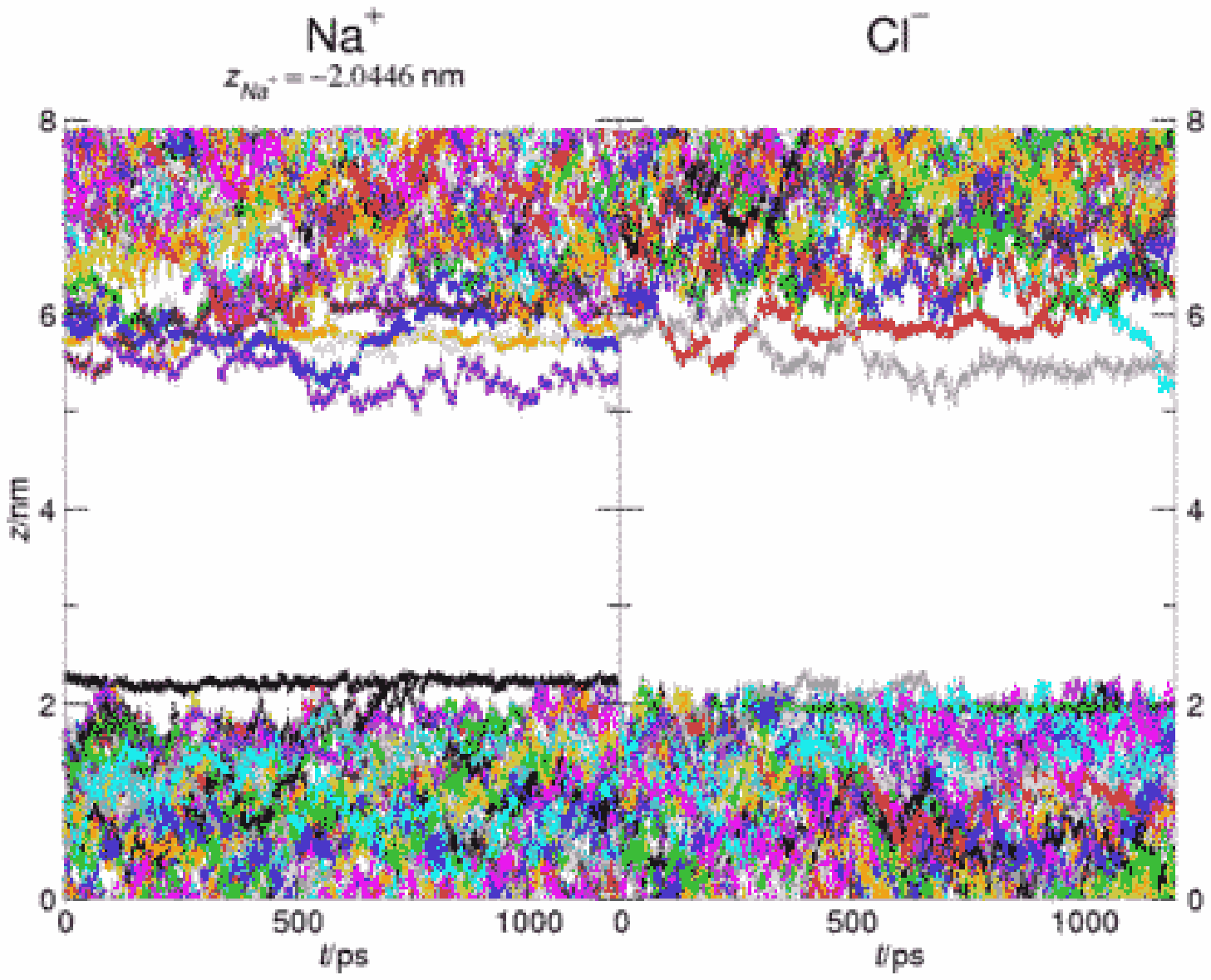}
  \includegraphics[width=\xw]{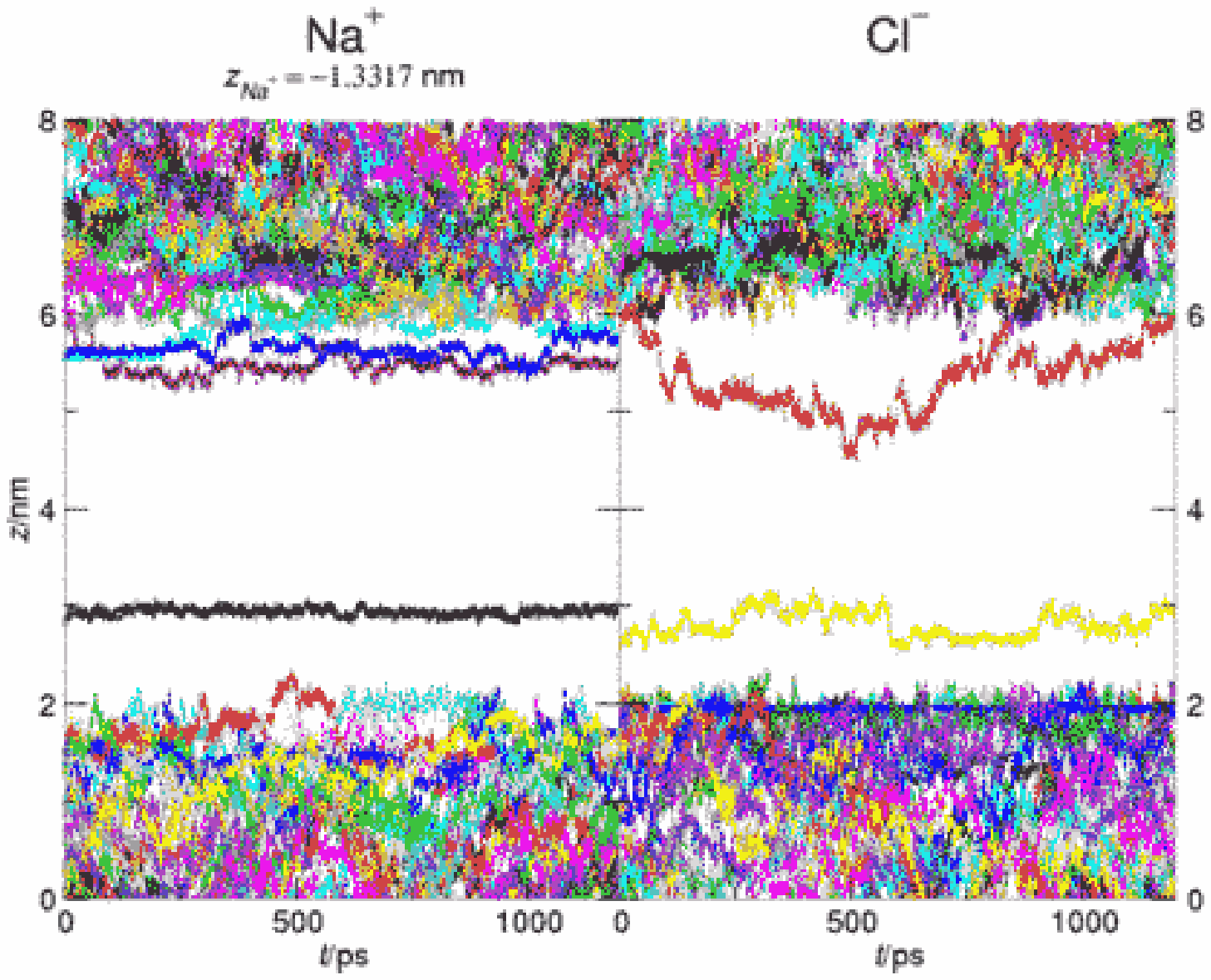}\\
  \includegraphics[width=\xw]{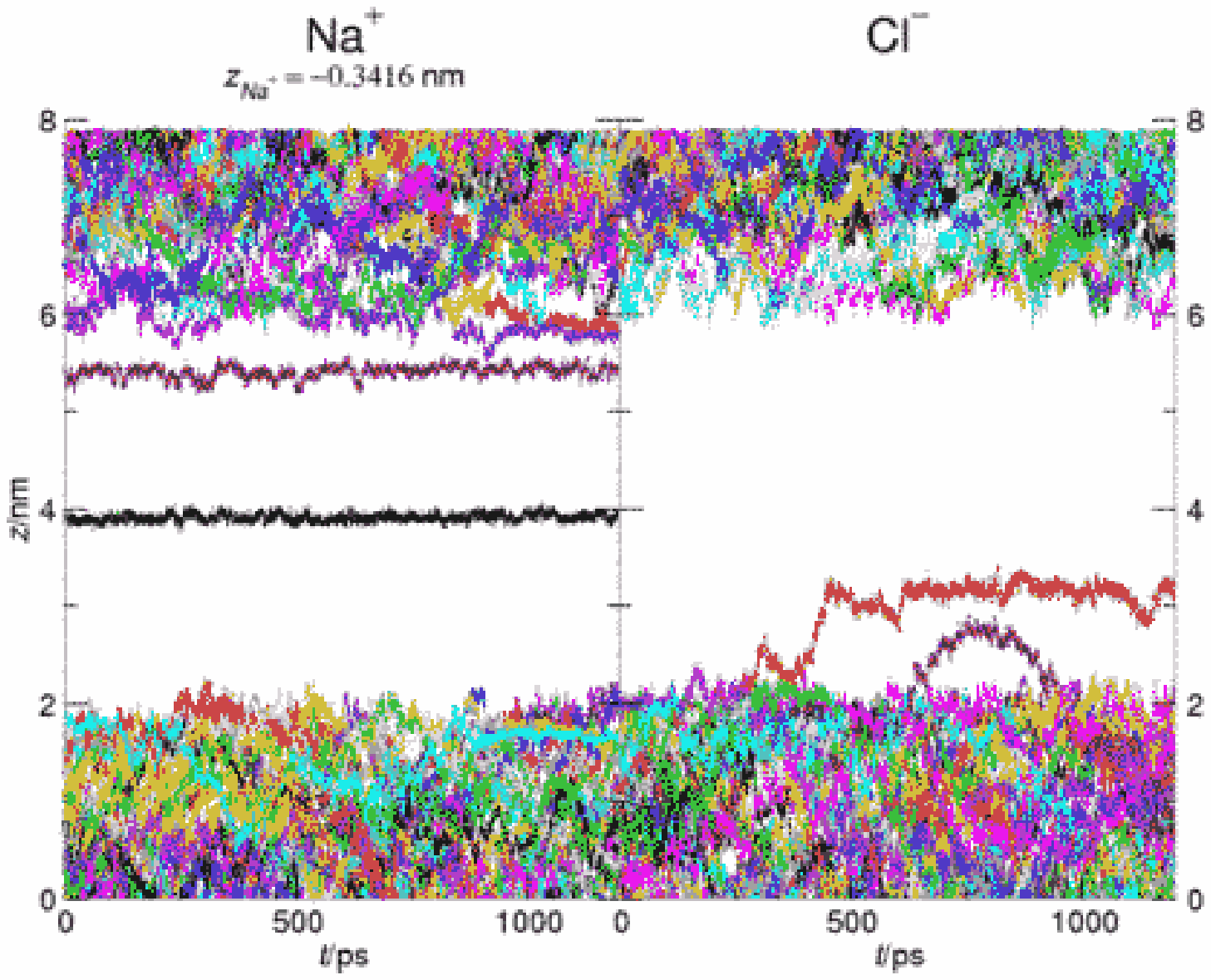}
  \includegraphics[width=\xw]{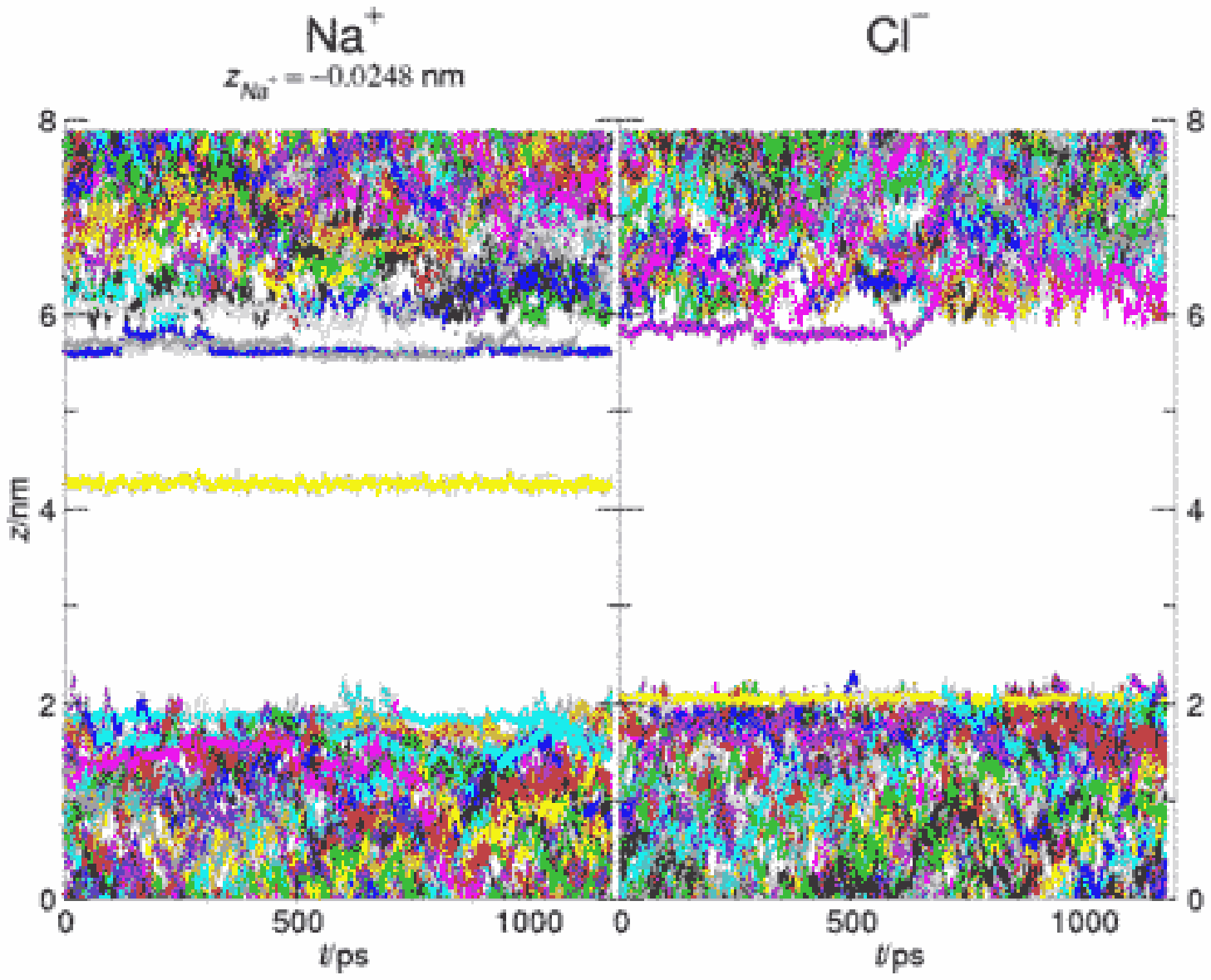}\\
  \includegraphics[width=\xw]{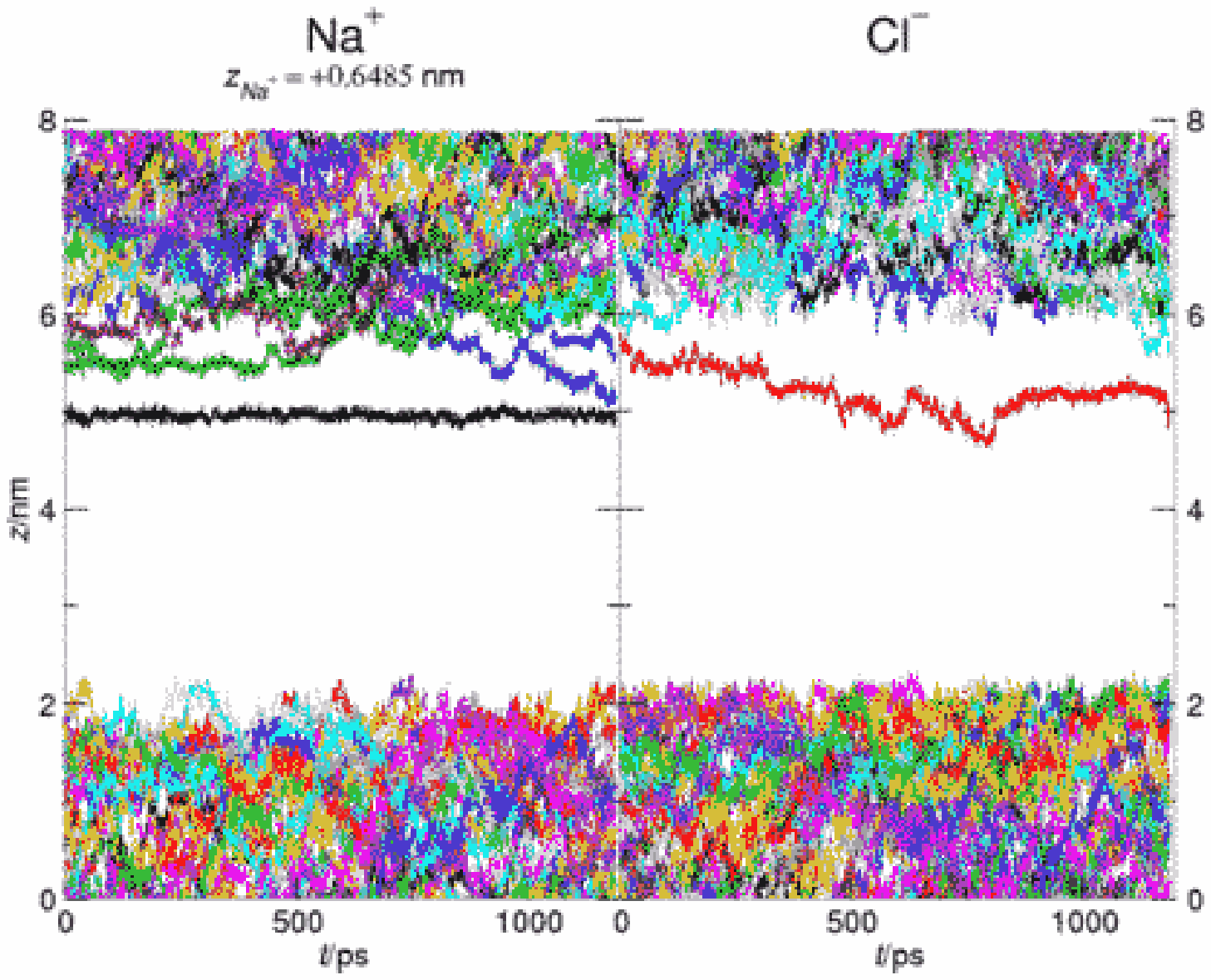}
  \includegraphics[width=\xw]{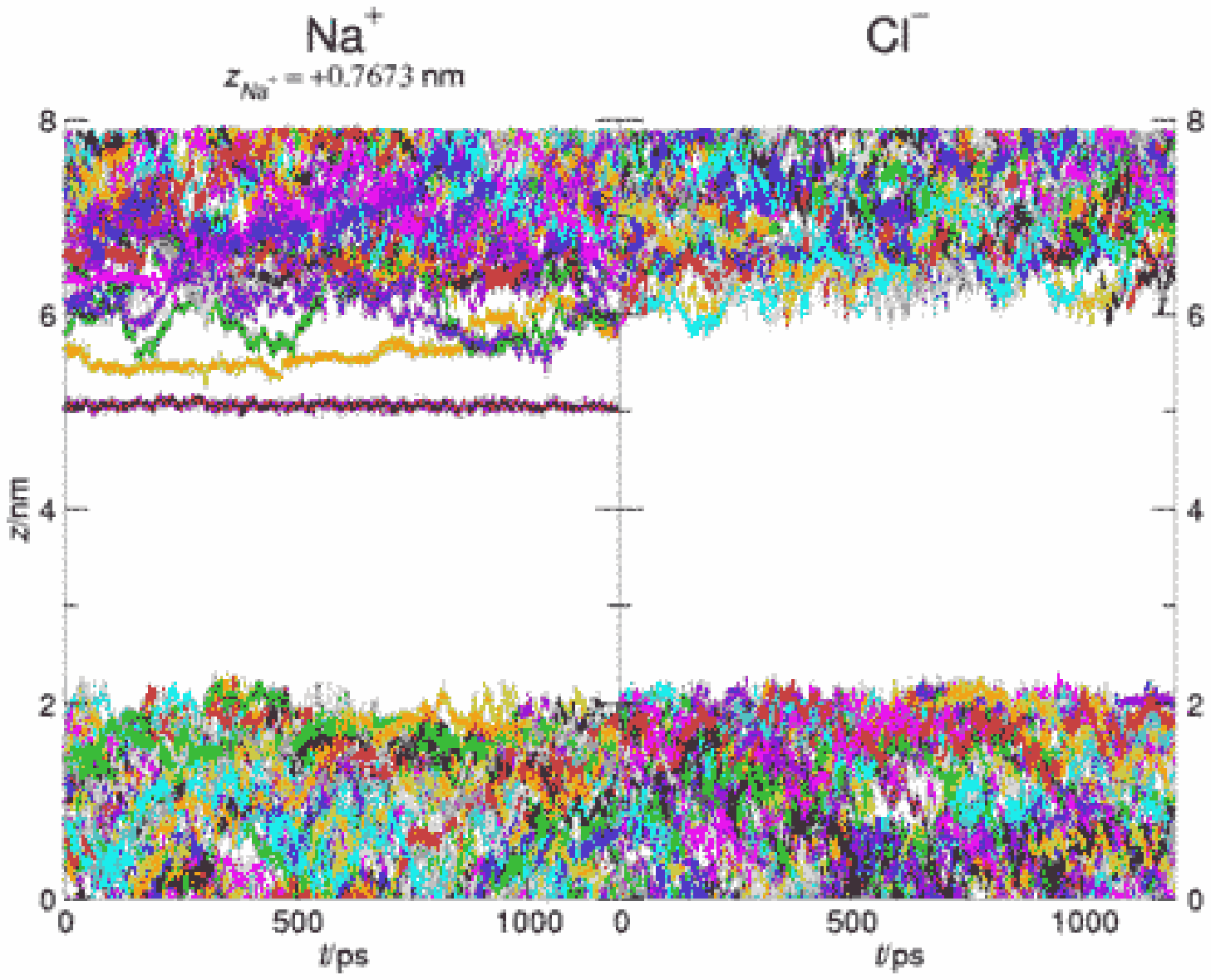}
  \caption{Position of all ions for selected umbrella
    windows. $z$-coordinates of the ions are given in absolute box
    coordinates whereas the main paper uses coordinates relative to
    the center of mass of the channel, $z_{0}=4.26$~nm. The one-ion
    pore region $\mathcal{P}_{1}$ was chosen as $2.1\ \text{nm} \leq z
    \leq 4.9\ \text{nm}$ (in box coordinates; in relative coordinates
    $-2.16\ \text{nm} \leq z-z_{0} \leq 0.64\ \text{nm}$). The
    umbrella-sampled \NA{} ion is always seen in the left panel as a
    oscillating line at the position indicated by the relative
    coordinate $z_{\text{\NA}}$. For instance, the windows for
    $z_{\text{\NA}}=-0.3416\ \text{nm}$, $z_{\text{\NA}}=-1.3317\
    \text{nm}$,  and $z_{\text{\NA}}=+0.6485\ 
    \text{nm}$ had to be rerun as described in the text to obtain
    trajectories with only the sampled ion in $\mathcal{P}_{1}$.}
  \label{fig:UMBzcoord}
\end{figure}

The trajectories show that the region $\mathcal{P}_{1}$ between
$-21.6$~\AA{} and $6.4$~\AA{} contains only the restrained \NA{} ion
for most of the time. Windows that did not meet this criterion were
rerun after the ions identified as entering $\mathcal{P}_{1}$ were
exchanged with water molecules in the bulk region. After this
procedure two windows still showed a significant density. Those were
simply rerun with a different initial distribution of velocities. Over
the time scale of the umbrella window this was sufficient to observe
no other ions entering the pore.

\setlength{\xw}{0.3\linewidth}
\begin{figure}[tbp]
  \centering
  \begin{minipage}{\linewidth}  
    \setlength{\unitlength}{0.01\linewidth}
    \begin{picture}(100,60)
      \put(2,0){\includegraphics[width=\xw]{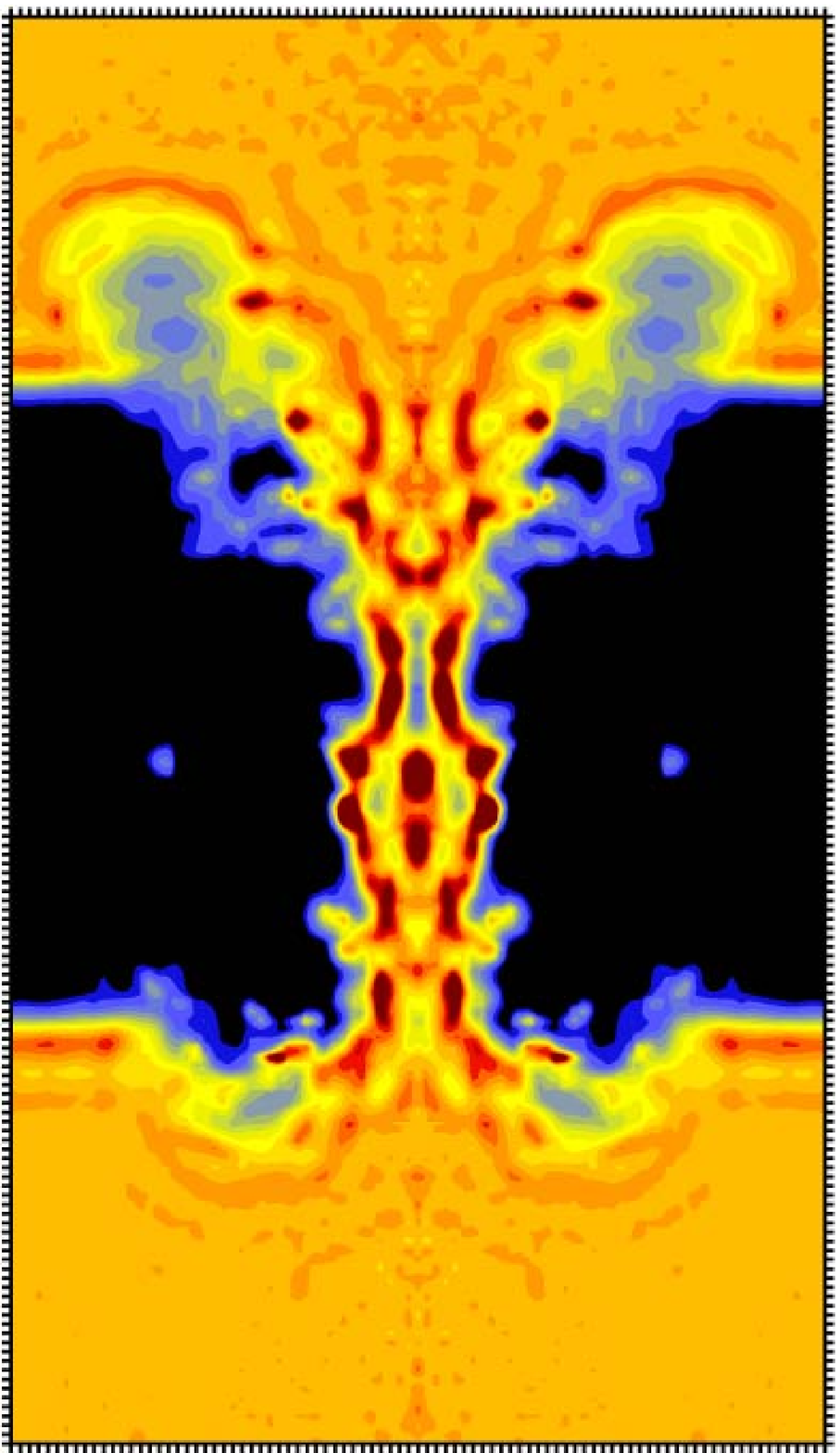}}
      \put(34,0){\includegraphics[width=\xw]{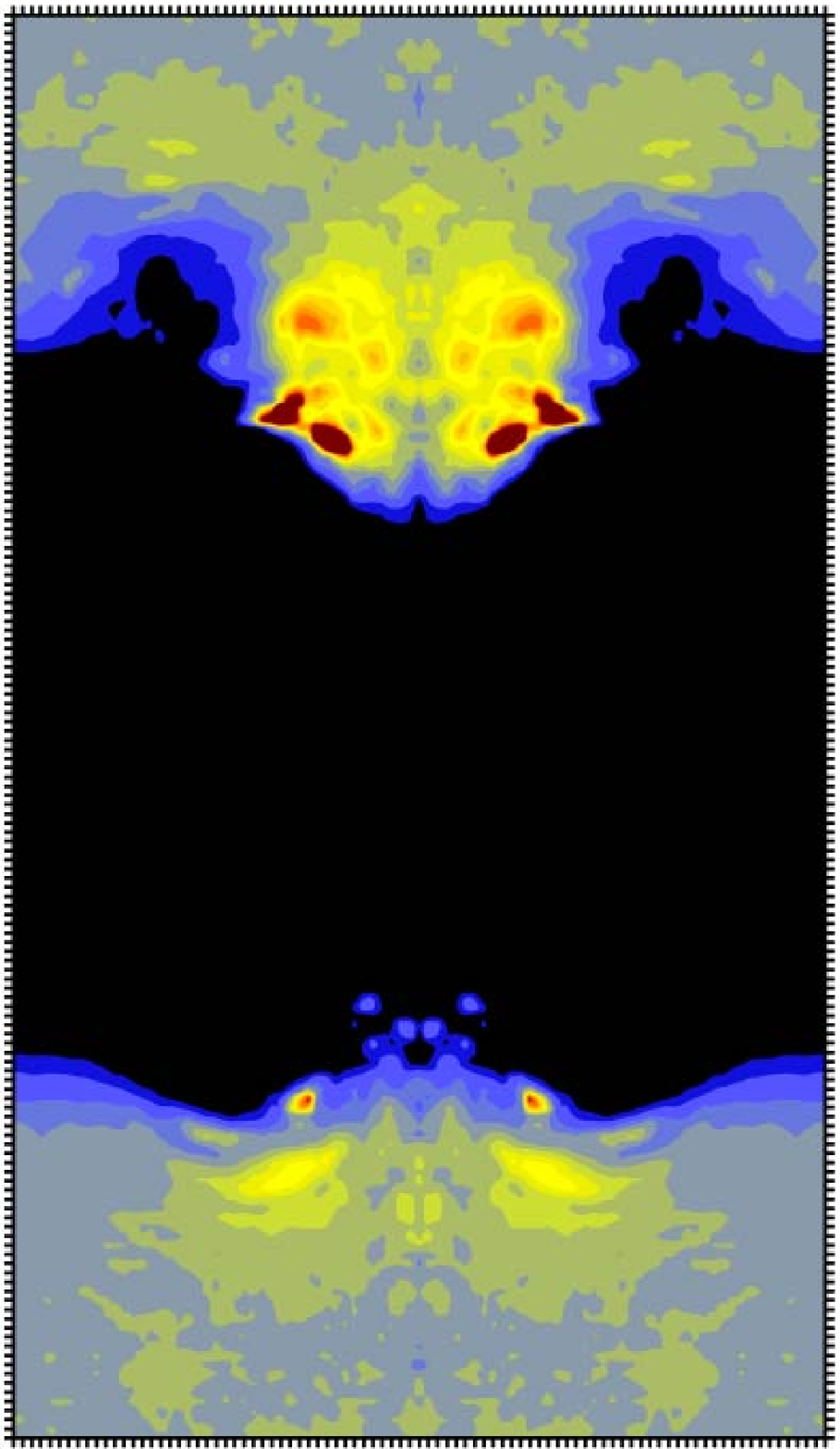}}
      \put(66,0){\includegraphics[width=\xw]{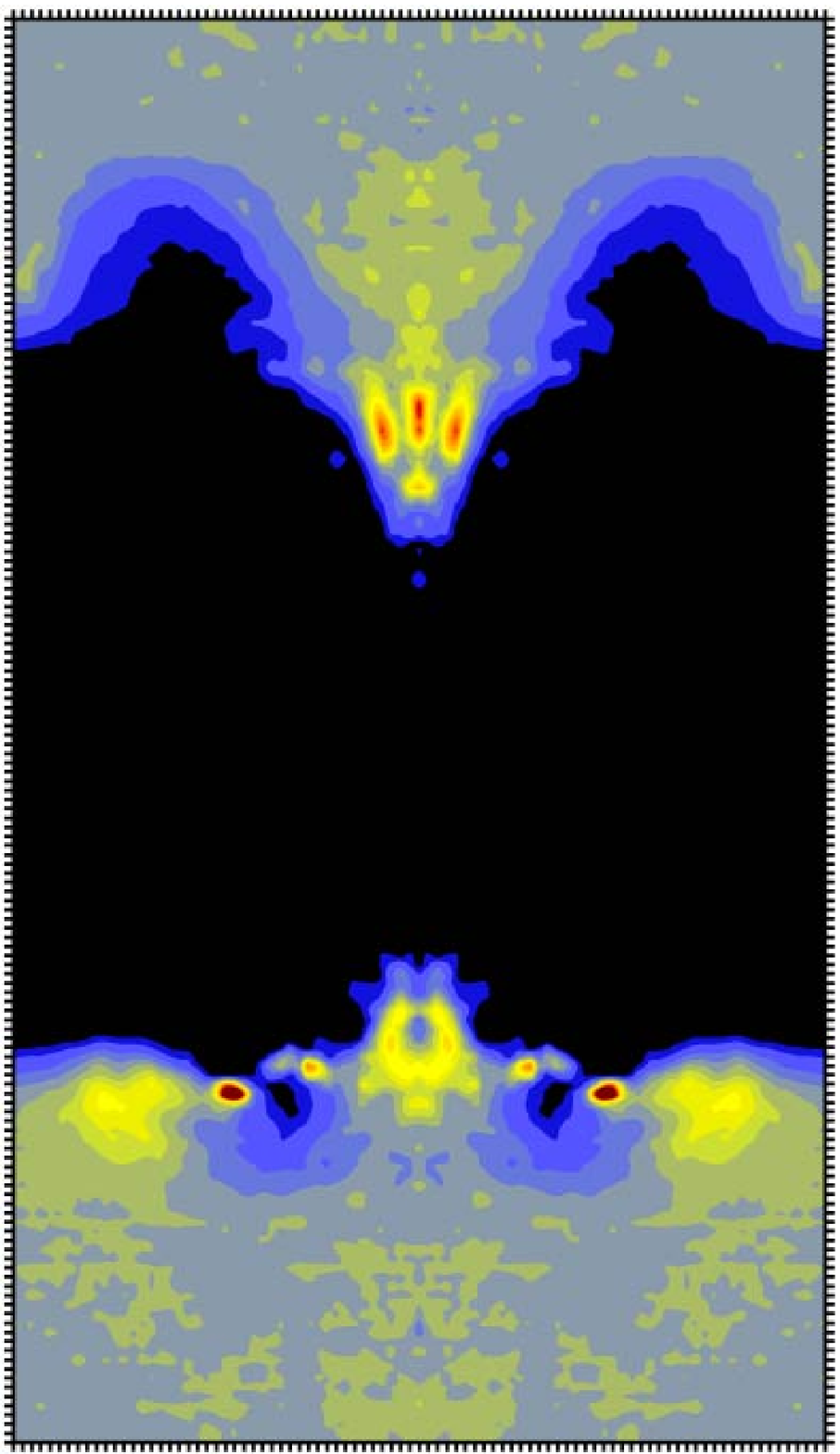}}
      \put(4,32){\makebox(0,0)[lb]
        {{\small\bf\sffamily\color[rgb]{1,1,1}\rule{0.82\linewidth}{1pt}1-ion}}}%
      \put(4,16){\makebox(0,0)[lb]
        {{\small\bf\sffamily\color[rgb]{1,1,1}\rule{0.82\linewidth}{1pt}1-ion}}}%
    \end{picture}
  \end{minipage}
  \caption{Density of water (left), \NA{} (center), and \CL{} (right),
    averaged over all umbrella windows from the one-ion region
    (between the two white lines). The sampled \NA{} ion was excluded
    from the \NA{} density. The density scales are the same as in
    figure~5 in the main paper (up to $1.5$ of the SPC bulk density
    for water and up to $4.5\ \text{mol}/\text{l}$ for ions).}
  \label{fig:UMBdensity}
\end{figure}

\begin{figure}[tbp]
  \centering
  \includegraphics[width=0.9\linewidth]{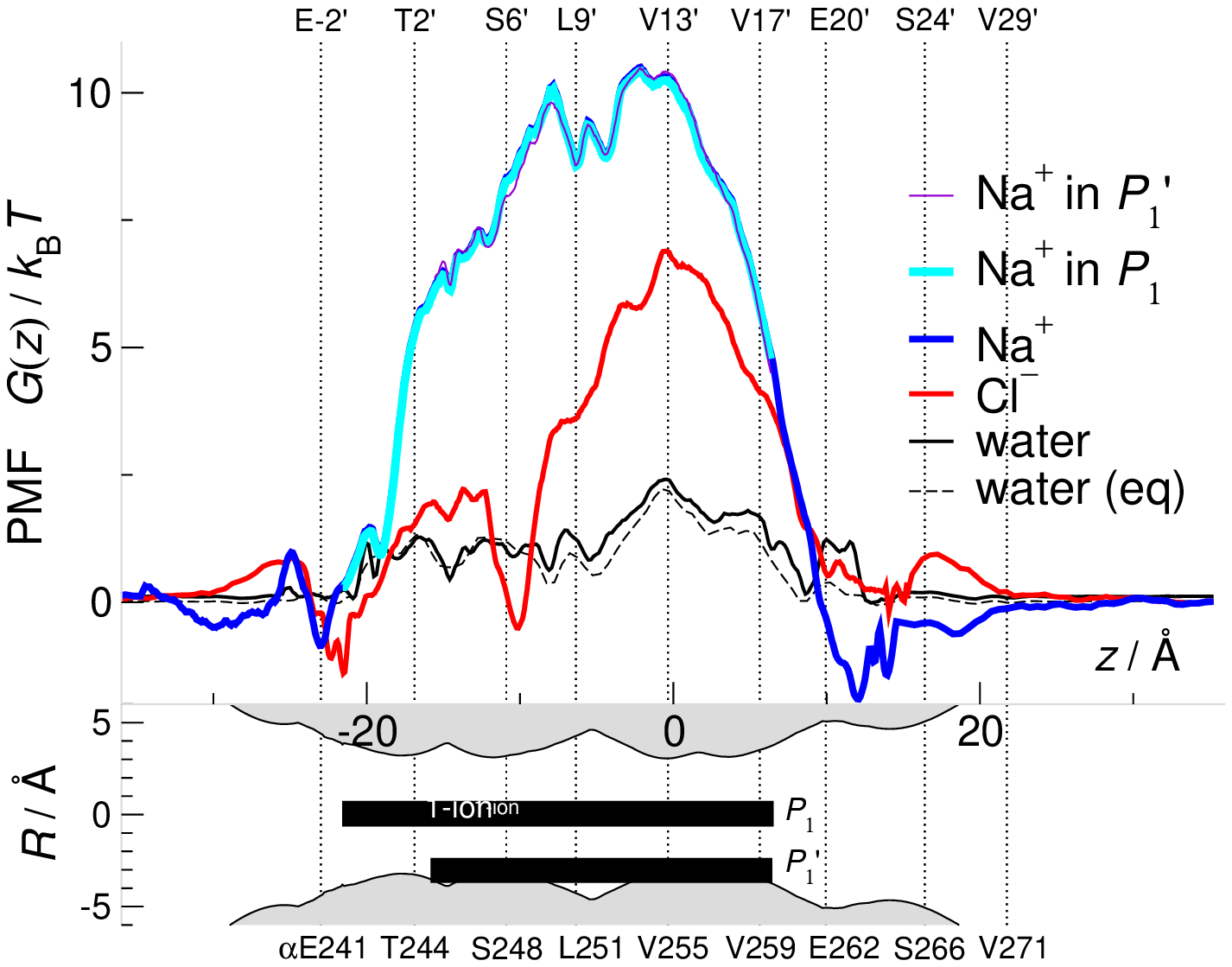}
  \caption{Potential of mean force. The \NA{} PMF (for the OPLS-AA
    force field) is computed over the region where there is strictly
    only the sampled ion in the pore ($\mathcal{P}'_{1}$), the region
    where this condition is fulfilled to a good approximation
    ($\mathcal{P}_{1}$) and the whole pore. The \CL{} PMF is calculated
    across the whole region, and so is the one for water. For
    comparison, the Boltzmann-sampled PMF for water is also shown.}
  \label{fig:pmf}
\end{figure}

Average densities from the umbrella windows,
figure~\ref{fig:UMBdensity}, show that $\mathcal{P}_{1}$ does not
contain any appreciable number of \NA{} ions though at the
intracellular entrance a few \CL{} ions venture across the boundary.
Based on this new evaluation one could argue that a more appropriate
one-ion region $\mathcal{P}'_{1}$ should be $-15.6\ \text{\AA} \leq z
\leq 6.4\ \text{\AA}$. However, calculating the \NA{} PMF over either
$\mathcal{P}'_{1}$ or $\mathcal{P}_{1}$ makes no appreciable
difference, as demonstrated in figure \ref{fig:pmf}. In addition, the
figure also shows that the one-ion PMF (whichever way calculated) does
not differ from the PMF calculated with contributions from regions
with multiple ion occupancies.

Lastly, figure~\ref{fig:pmf} shows that the Boltzmann-sampled and
umbrella-sampled PMF for water are close to identical (perhaps with a
small exception near E$20'$). This indicates that at least both
multi-particle PMFs are converged to the same degree. Only in the
mouth regions can we compare Boltzmann-sampled and umbrella sampled
ion PMFs. Generally, the agreement is good (especially when using a
harmonic confinement potential for umbrella sampling), which seems to
indicate that we can use the ion PMFs in the same way as the water
PMF. For the umbrella sampled \NA{} PMF we carried out a bootstrap
analysis for WHAM. The unbiasing procedure was repeated 1000 times
while the input data was drawn randomly with repetition from the
original data set. The resulting error bars are too small to show up
in figure~\ref{fig:pmf} and are $<0.05\,kT$. Thus, the data are not
heterogeneous and sufficient to lead to well defined PMFs. Lastly, the
tolerance of the WHAM procedure was varied between $10^{-3}\,kT$ and
$10^{-7}\,kT$. The PMFs did not change beyond $10^{-5}\,kT$ so this
value was chosen throughout.

For WHAM analysis we use Alan Grossfield's \texttt{wham} code
(\url{http://dasher.wustl.edu/alan/wham/index.html}), modified for use
with Gromacs output files.  An advantage of using umbrella sampling
and WHAM is that one can always add more windows to a data set to
improve convergence. The force constant for those simulations does not
have to be the same as for the initial windows because the unbiasing
procedure takes care of this. We took advantage of this fact so that
the actual umbrella sampling parameters could vary within a system and
also from system to system.  They are listed in
table~\ref{tab:umbparams}. 

The nicotinic receptor is a cation selective channel and thus
investigating gating was necessarily an investigation of how the \NA{}
current was influenced by the protein structure. Thus, the most
important calculations were the ones for the \NA{} PMF in the OPLS-AA
force field. Depending on the hardware, 1~ns of umbrella sampling took
between 11h (using four processors on a quad Opteron cluster) and 28h
(with two processors on a dual Pentium IV cluster).  Using more
processors across nodes was not effective as the relatively small
system size (ca 15,000 atoms) leads to a large communications overhead
and rapidly diminishing returns.

\section{Water in the hydrophobic constriction}

Both the water PMF (figure~\ref{fig:pmf}) and the water density
(figure~\ref{fig:UMBdensity}) unanimously show that the whole nAChR
pore is water filled. These two quantities determine the equilibrium
properties of a system (and the latter is a true equilibrium average).
As such they do not contain explicit information about fluctuations
such as strong density fluctuations (`liquid-vapor oscillations'
\citep{Beckstein03}) in the pore. For MscS, \citet{Anishkin04}
reported MD simulations during which the gate was predominantly empty
(vapor filled), with some intermittent water filling.  They put
forward a `vapor-lock' hypothesis, stating that the absence of water
will exclude the passage of ions in MscS and perhaps nAChR. This is
certainly a physically sound hypothesis, especially when results from
simulations in hydrophobic model pores are taken into account
\citep{Beckstein01,Allen02,Allen03b,Beckstein03,Beckstein04,Beckstein04a}.
Hence it is worthwhile investigating the possibility of a vapor state
in the cryo-EM model of the nAChR pore.

\setlength{\xw}{0.3\linewidth}
\begin{figure}[tbp]
  \centering
  \includegraphics[width=\xw]{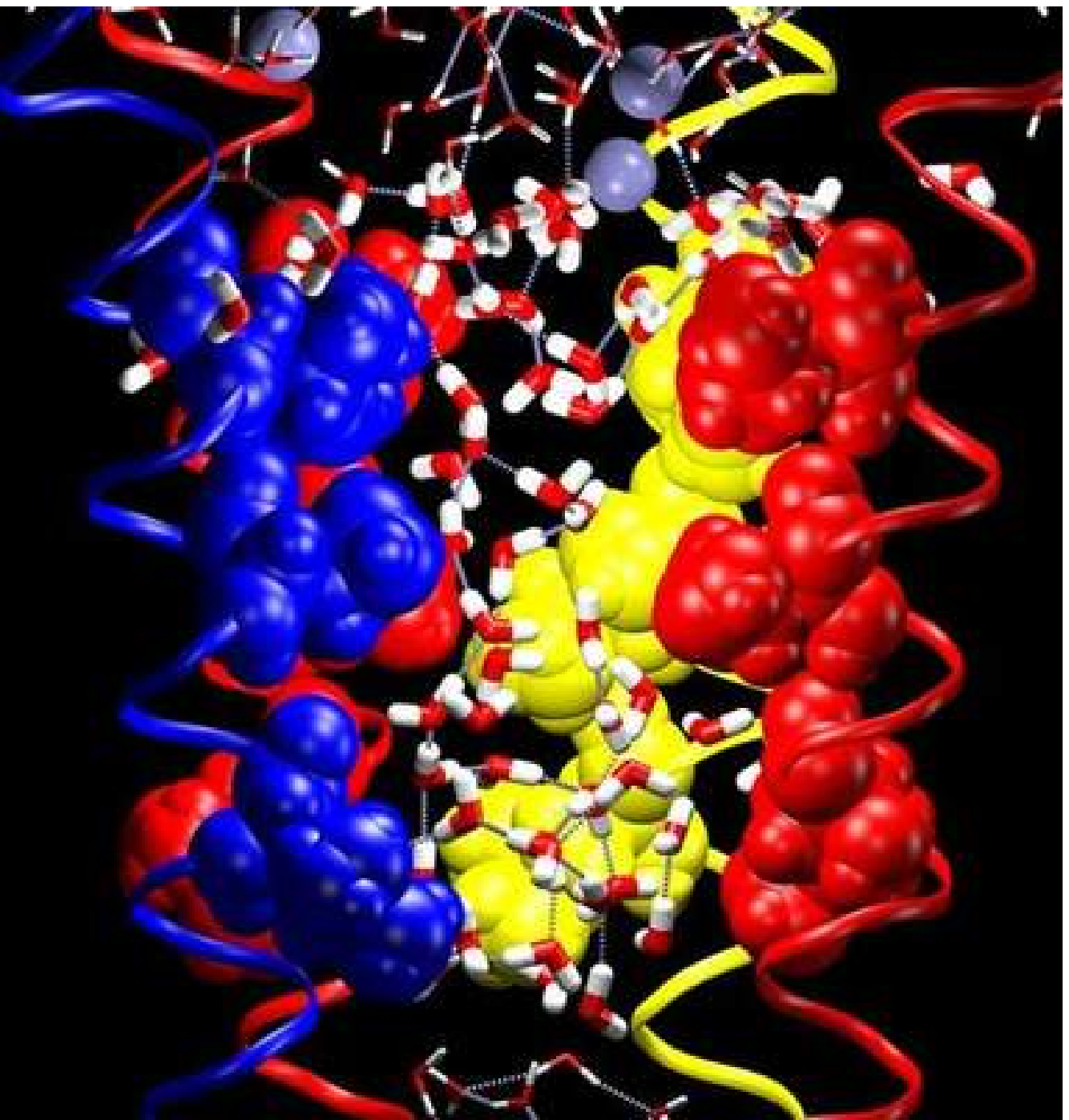}
  \includegraphics[width=\xw]{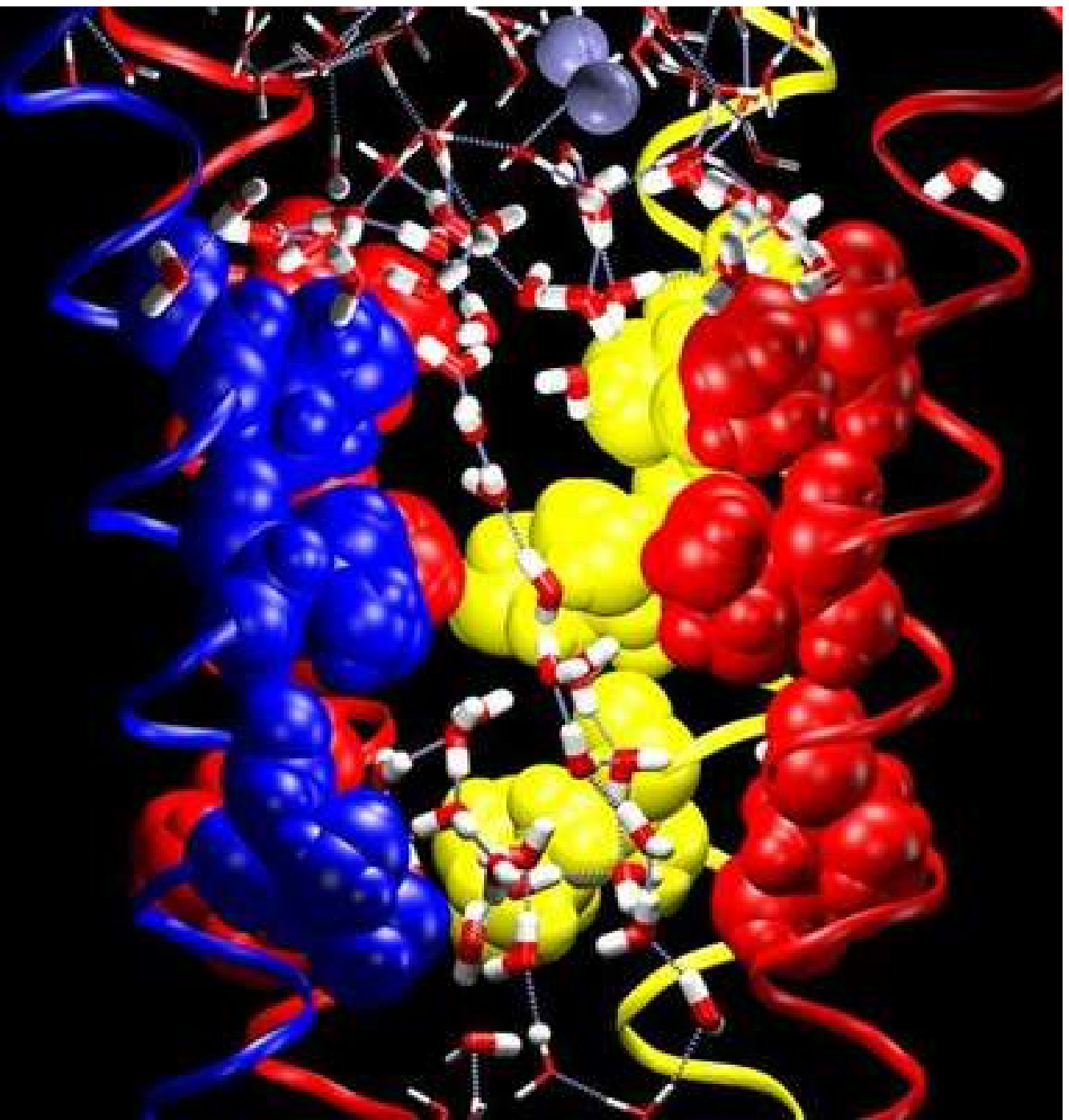}
  \includegraphics[width=\xw]{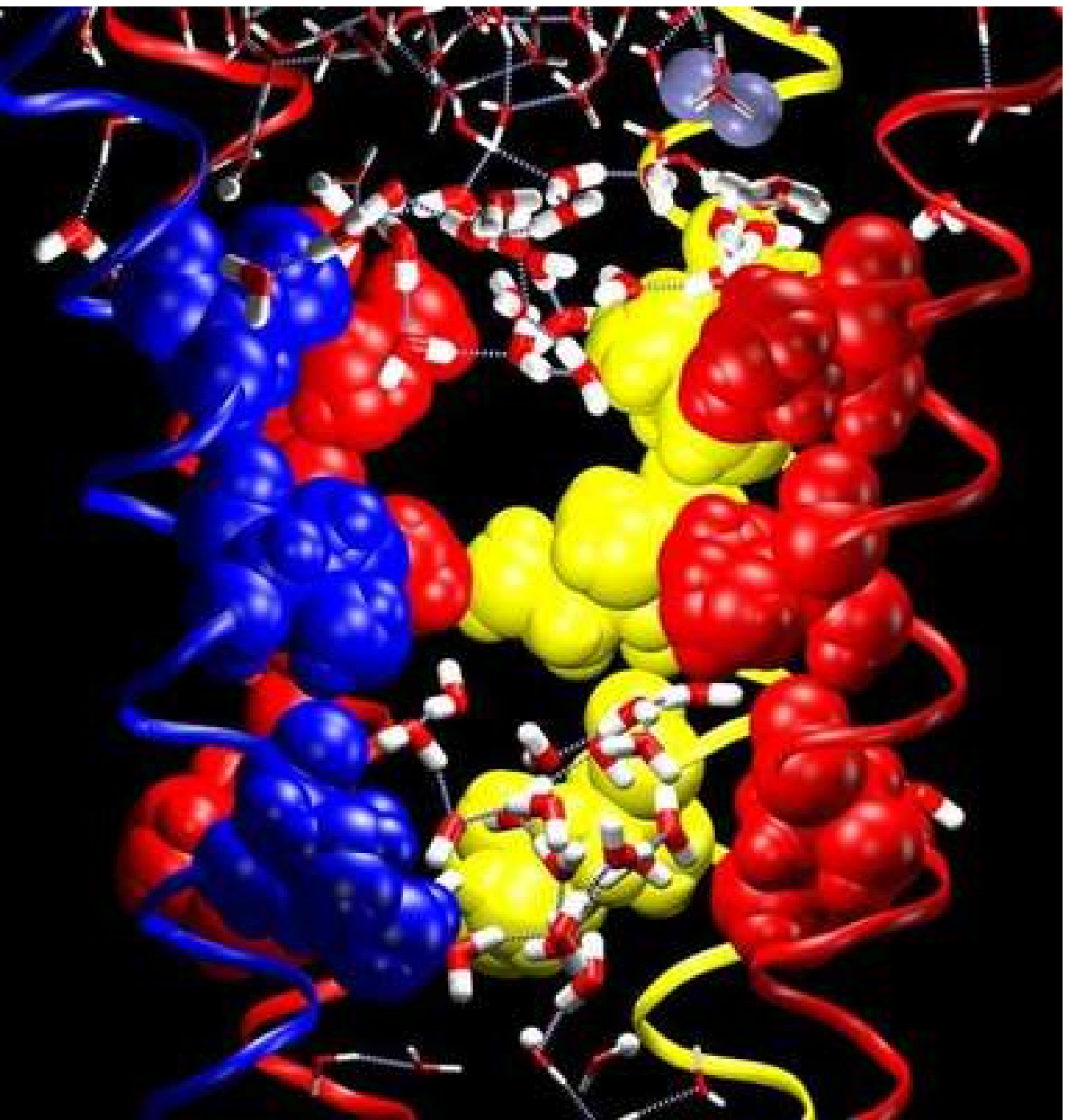}
  \caption{Behavior of water in the hydrophobic constriction at the
    $13'$ position. Hydrogen bonds are indicated by blue broken
    lines. The hydrophobic girdle residues at $9'$, $13'$ and $17'$ are
    shown in van der Waals representation and colored by subunit (red:
    $\alpha$, yellow: $\gamma$, blue: $\delta$). The $\beta$ subunit
    is omitted to provide an unobstructed view on the pore. \NA{} ions
    are shown in light blue.} 
  \label{fig:watergate}
\end{figure}

Analysis of the 60~ns equilibrium simulation shows that there is only
a short region of about $4.2$~\AA{} length around the $13'$
hydrophobic ring (see table~\ref{tab:girdle} for those residues) that
ever switches to a vapor phase. In the main paper we shown that the
vapor phase only accounts for 19\% of the total simulation time. Thus,
a vapor state is unlikely to be the sole reason for a barrier to ion
permeation. Nevertheless, it is interesting to investigate the
behavior of water in the hydrophobic constriction more closely.

\begin{table}[!b]
  \centering
  \caption{Residues in the hydrophobic girdle of the muscle-type
    nicotinic acetylcholine receptor from \textit{Torpedo} as seen in
    the pdb structures 1OED \protect\citep{Miyazawa03} and 2BG9
    \protect\citep{Unwin05}.} 
  \label{tab:girdle}
  \begin{tabular}{cccccc}
    \toprule
    position & $\alpha_{\delta}$ & $\gamma$ & $\alpha_{\gamma}$
             & $\beta$  & $\delta$\\
    \midrule
    $17'$ & V259 & L271 & V259 & L266 & L273\\
    $13'$ & V255 & I267 & V255 & V261 & V269\\
     $9'$ & L251 & L263 & L251 & L257 & L265\\
    \bottomrule
  \end{tabular}
\end{table}
 
In figure~\ref{fig:watergate} we show typical snapshots from the
trajectory (a short movie is also available at
\url{http://sbcb.bioch.ox.ac.uk/oliver/download/Movies/watergate.mpg}).
The most common state of the pore is filled by water as shown in the
first snapshot, without any obvious ordering (apart from the
preference for the five positions seen in the density and the
avoidance of the center of the pore).  Occasionally a string of water
molecules forms across the constriction site (see the second
snapshot). Finally, the constriction can be void of water although for
much shorter periods than what we previously observed in hydrophobic
model pores of comparable radii \citep{Beckstein03}.

\newlength{\figH}
\begin{figure}[tb]
  \setlength{\figH}{0.3\textheight}
  \centering
  \subfigure[\mbox{PMF} $\Delta G(z)$]%
  {\label{fig:ionpmf-wLpmf-pmf}%
    \includegraphics[clip,height=\figH]{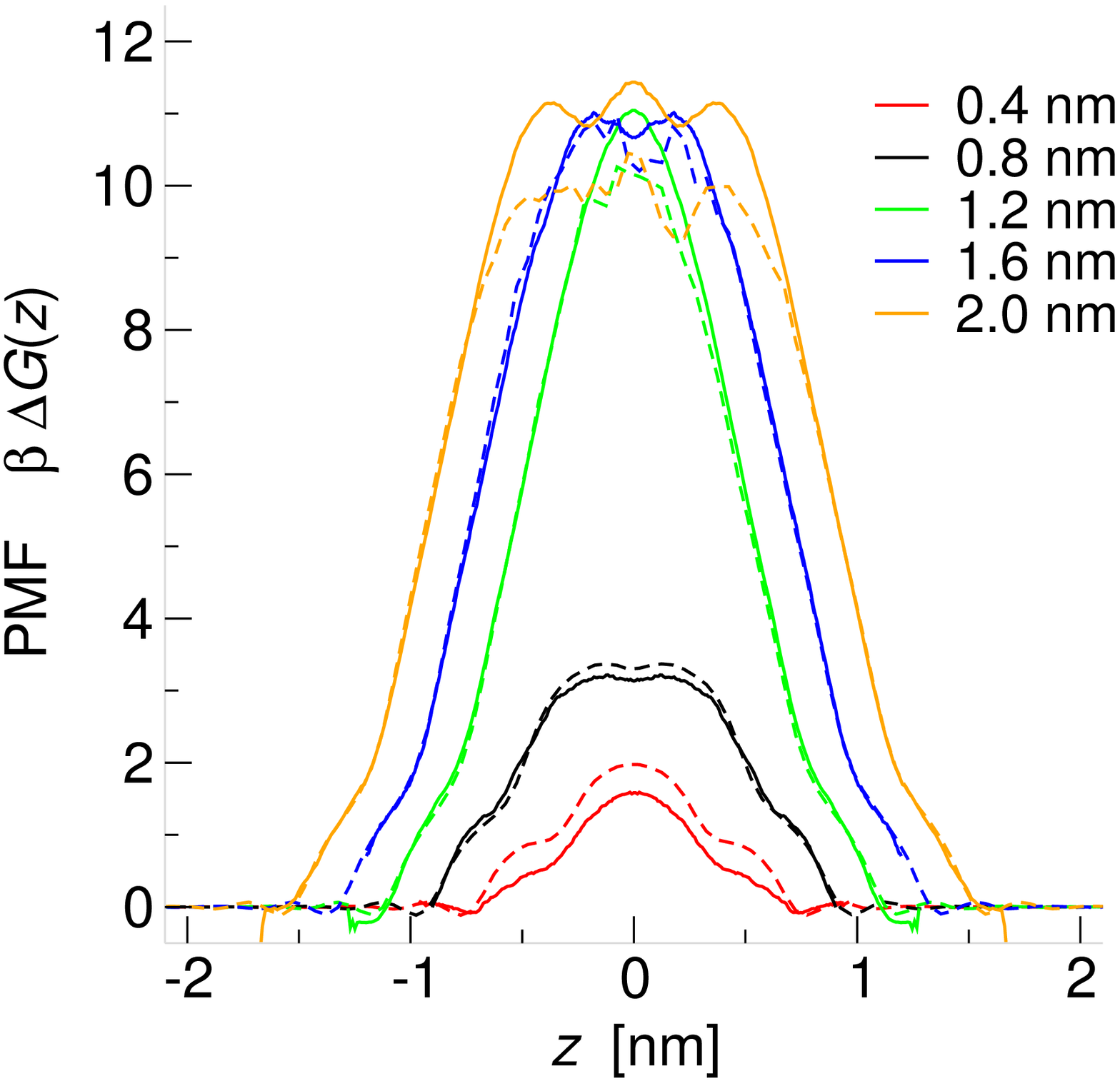}}%
  \hfill%
  \subfigure[\mbox{Barrier} height $\Delta G^{\ddagger}$]%
  {\label{fig:ionpmf-wLpmf-barrier}%
    \includegraphics[clip,height=\figH]{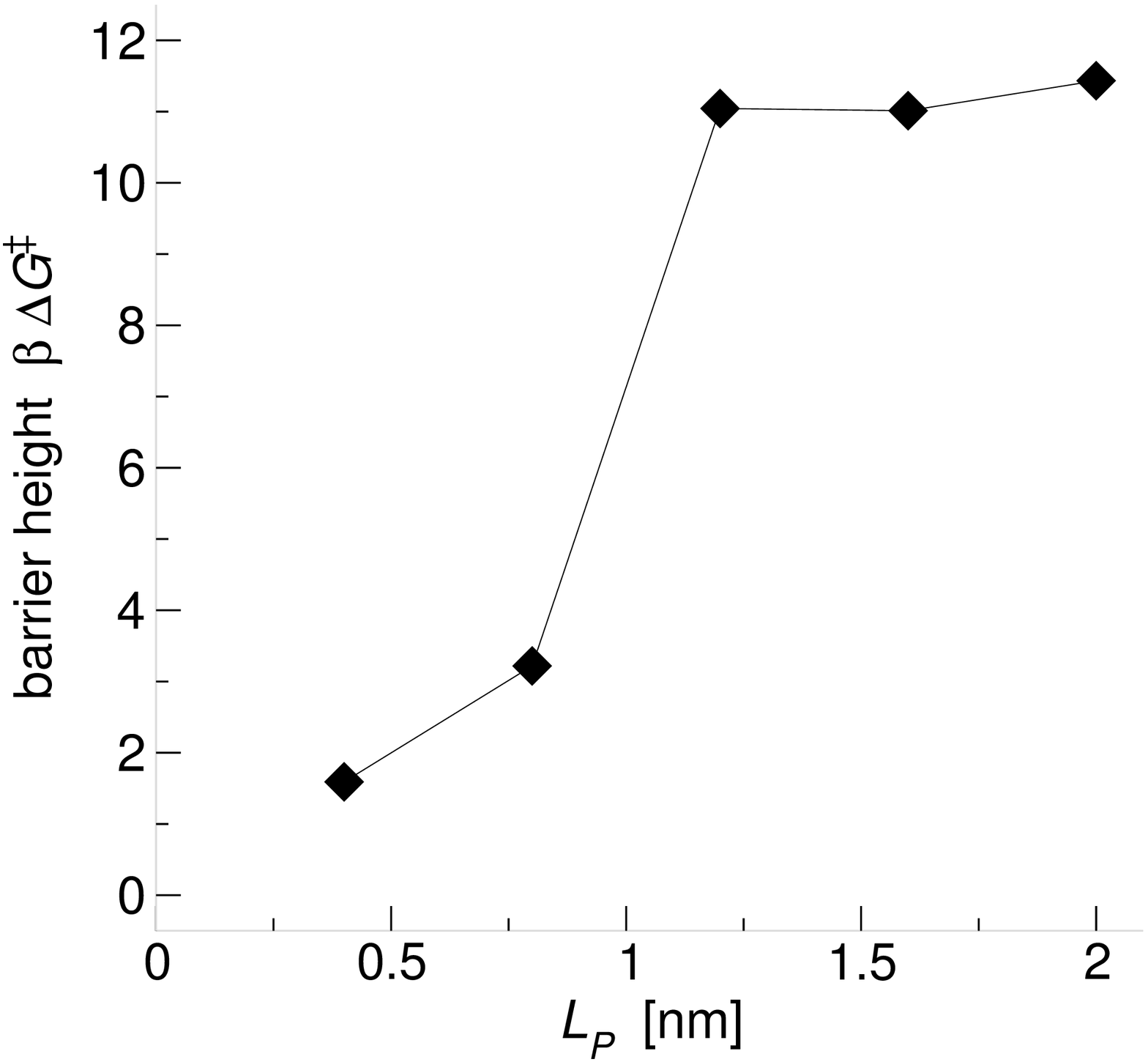}}
  \caption[\mbox{PMF} of a water molecule in pores of differing length]%
  {\mbox{Umbrella-sampled} PMF of a water molecule in pores of
    differing length with a fixed pore radius of $R=0.55$~nm. The
    umbrella PMF is shown as a solid line, with the equilibrium
    density-derived PMF in broken style. Umbrella sampled PMFs were
    symmetrized. (Figure reproduced from \citep{Beckstein04T})}
  \label{fig:ionpmf-wLpmf}
\end{figure}

However, an important difference between the nAChR constriction and
the previous work is the \emph{length} of the pore. Those previous
calculations were carried out for pores of $L=8$~\AA{} whereas the
simulations show that the hydrophobic nAChR constriction has a length
less than $4.2$~\AA.  Our unpublished work \citep{Beckstein04T} shows
a very strong dependence of the water behavior on the pore length for
$L < 12$~\AA{}: as seen in figure~\ref{fig:ionpmf-wLpmf} the PMF for a
water molecule in a hydrophobic pore (albeit with radius
$R=5.5$~\AA{}) changes considerably between $L=4$~\AA{} and
$L=12$~\AA.

It appears that the hydrophobic constriction behaves as a hydrophobic
nano pore with $L=4$~\AA{} and $R=3$~\AA. Using the simple
thermodynamic model from \citep{Beckstein04} we can compute the
probability for the pore to be in the liquid filled state,
\begin{equation}
  \label{eq:FEtoOpenness}
  p_{\text{liq}} = \frac{1}{1+\exp[-\beta\,\Delta\Omega(R,L,\theta_{e})]},
\end{equation}
(equation~6 in \citep{Beckstein04}) from the free energy difference
(equation~5 in \citep{Beckstein04})
\begin{equation}
  \label{eq:DOmegaTheta}
  \Delta\Omega(R,L,\theta_{e}) = 2\pi\,R\,\gamma_{lv}(R + L\,\cos\theta_{e}).
\end{equation}
If we simply use the parameters extracted from the previous
simulations of hydrophobic pores (table~1 in \citep{Beckstein04}),
namely the water surface tension $\gamma_{lv} =
10\,kT\,\text{nm}^{-2}$ and the hydrophobicity of the pore (the
contact angle) $\theta_{e}=134^{\circ}$ we arrive at $\Delta\Omega =
0.42 \,kT$ and $p_{\text{liq}}=0.60$ or a probability for the vapor
state of 40\%. This number has the same order of magnitude as the 19\%
extracted from the simulations---rather good agreement for such a
crude model. Of course, this is not a rigorous derivation (and it is
fairly sensitive to the geometric dimensions) but it seems to indicate
that we can apply a simplified model to the well defined hydrophobic
constriction of nAChR. (It is amusing to note that our initial nano
pores \citep{Beckstein01} were designed to mimic the dimensions of the
hydrophobic girdle, based on the low resolution structure
\citep{Unw93,Unw00}, two years \emph{before} the near atomic
resolution structure was published \citep{Miyazawa03}. Now the MD
simulations show that we were justified in choosing this simplified
model but the local environment of the hydrophobic girdle really
admits water to a greater degree so that the length of the model pores
should have been closer to $4$~\AA.)

The appearance of a vapor state is an indicator of the hydrophobic
nature of the pore. If the pore was extremely hydrophobic then the
vapor state would be the stable thermodynamic state and it would
certainly block ion flow \citep{Beckstein04}. A desolvation barrier
for ions only requires moderately hydrophobic confinement
\citep{Beckstein04a}, so the absence of vapor does not necessarily
mean that a channel is open. As with many phenomena in biology it
turns out that evolution has finely shaped and balanced the
physical-chemical forces to make best use of them.

%
%
\newcommand{\newblock}{\relax}


%